\documentclass[conference]{IEEEtran}
\usepackage{xspace}
\usepackage{pifont}
\usepackage{algorithm}
\usepackage{enumitem}
\usepackage{makecell}
\usepackage[noend]{algpseudocode}
\usepackage[normalem]{ulem}
\usepackage{subfigure}
\usepackage{graphicx}
\usepackage{amsmath}

\usepackage{amssymb}
\usepackage{totpages}
\usepackage{booktabs}
\usepackage{comment}
\usepackage{authblk}
\usepackage[capitalise]{cleveref}

\numberwithin{equation}{section}

\newcommand{\name}{\textsc{LOFT}\xspace}
\newcommand{\mcname}{multi-class \name}
\newcommand{\nameacr}{Low-Rate Overuse Flow Tracer\xspace}
\newcommand{\fpart}{update algorithm\xspace}
\newcommand{\spart}{estimation algorithm\xspace}

\newcommand{\etal}{et~al.\@\xspace}

\renewcommand{\paragraph}[1]{\noindent\textbf{#1}.}

\newcommand{\estim}{U}
\newcommand{\accum}{A}
\newcommand{\counter}{C}
\newcommand{\flowseg}{X}
\newcommand{\fastfreq}{\omega}
\newcommand{\rcounter}{c}
\newcommand{\parts}{\theta}
\newcommand{\flowsize}{F}
\newcommand{\llabel}{l}
\newcommand{\blabel}{b}
\newcommand{\midp}{\tau}
\newcommand{\flownum}{N}
\newcommand{\boundv}{M}
\newcommand{\fsgamma}{\gamma}
\newcommand{\fsbeta}{\beta}
\newcommand{\parttime}{\fastfreq^{-1}}
\newcommand{\slowfreq}{Z}
\newcommand{\fwidth}{W}
\newcommand{\fmwidth}{W_{\mathit{fm}}}
\newcommand{\dtime}{t}
\newcommand{\ceil}[1]{\lceil #1 \rceil}
\newcommand{\ratio}{\ell}
\newcommand{\boundf}{B}
\newcommand{\resetcycle}{\parts_{reset}}

\newif\ifremovecomments
\removecommentsfalse

\makeatletter
\newcommand{\mysubscript}[2]{\textsubscript{\textcolor{#2}{\textsf{\textbf{#1}}}}}
\newcommand*\defcomment[4]{
  \ifremovecomments
    \expandafter\newcommand\csname #1\endcsname[1]{%
    }
    \expandafter\newcommand\csname @#2delnoname\endcsname[1]{%
    }
    \expandafter\newcommand\csname #2del\endcsname[1]{%
    }
    \expandafter\newcommand\csname #2sugg\endcsname[1]{##1}
    \expandafter\newcommand\csname #2subs\endcsname[2]{##2}
  \else
    \expandafter\newcommand\csname #1\endcsname[1]{%
      \textcolor{#4}{\ding{110}\mysubscript{#3}{#4}\,{##1}}%
    }
    \expandafter\newcommand\csname @#2delnoname\endcsname[1]{%
      \bgroup\markoverwith{\textcolor{#4}{\rule[0.35ex]{2pt}{1pt}}}\ULon{##1}%
    }
    \expandafter\newcommand\csname #2del\endcsname[1]{%
      \csname @#2delnoname\endcsname{##1}\kern0.1em\mysubscript{#3}{#4}%
    }
    \expandafter\newcommand\csname #2sugg\endcsname[1]{%
      \textcolor{#4}{[##1]\mysubscript{#3}{#4}}%
    }
    \expandafter\newcommand\csname #2subs\endcsname[2]{%
      \csname @#2delnoname\endcsname{##1}\csname #2sugg\endcsname{##2}%
    }
  \fi
  \expandafter\newcommand\csname #2sout\endcsname{\csname #2del\endcsname}
}
\makeatother
\defcomment{daniele}{da}{DA}{magenta}
\defcomment{adrian}{ap}{AP}{blue}
\defcomment{hc}{hc}{HC}{red}
\defcomment{yuhsi}{yc}{YC}{violet}
\defcomment{che}{che}{CY}{orange}
\defcomment{simon}{ss}{SS}{brown}
\defcomment{ben}{br}{BR}{magenta}
\defcomment{ml}{ml}{ML}{green}

\usepackage{siunitx}
\DeclareSIUnit{\ns}{\nano\second}
\DeclareSIUnit{\ms}{\milli\second}
\DeclareSIUnit{\us}{\micro\second}

\DeclareSIUnit{\B}{\byte}
\DeclareSIUnit{\kB}{\kilo\byte}
\DeclareSIUnit{\KB}{\kB}
\DeclareSIUnit{\MB}{\mega\byte}
\DeclareSIUnit{\GB}{\giga\byte}

\DeclareSIUnit{\mbps}{Mbps}
\DeclareSIUnit{\gbps}{Gbps}
\DeclareSIUnit{\tbps}{Tbps}
\DeclareSIUnit{\mpps}{Mpps}

\begin{document}


\author[1]{Simon Scherrer}
\author[2]{Che-Yu Wu}
\author[2]{Yu-Hsi Chiang}
\author[1]{Benjamin Rothenberger}
\author[1]{Daniele E. Asoni}
\author[3]{\authorcr Arish Sateesan}
\author[3]{Jo Vliegen}
\author[3]{Nele Mentens}
\author[2]{Hsu-Chun Hsiao}
\author[1]{Adrian Perrig}
\affil[1]{Department of Computer Science, ETH Zurich, Switzerland \authorcr Email: { \{simon.scherrer,benjamin.rothenberger,daniele.asoni,adrian.perrig\}@inf.ethz.ch}}
\affil[2]{National Taiwan University, Taiwan\authorcr Email: \{r06922021,r06922023,hchsiao\}@ntu.edu.tw}
\affil[3]{ESAT, KU Leuven, Belgium \authorcr Email: \{arish.sateesan,jo.vliegen,nele.mentens\}@kuleuven.be}

\title{\nameacr (\name):\\An Efficient and Scalable Algorithm \\for Detecting Overuse Flows}
\maketitle

\thispagestyle{plain} 
\pagestyle{plain}

\begin{abstract}
	Current probabilistic flow-size monitoring can only detect heavy hitters
(e.g., flows utilizing \num{10} times their permitted bandwidth), but
cannot detect smaller overuse (e.g., flows utilizing \numrange{50}{100}\% more
than their permitted bandwidth).
Thus, these systems lack accuracy in the challenging
environment of high-throughput packet processing, where fast-memory resources are scarce. 
Nevertheless, many applications rely
on accurate flow-size estimation, e.g. for network monitoring, anomaly detection
and Quality of Service.

We design, analyze, implement, and evaluate \name, a new approach for
efficiently detecting overuse flows that achieves dramatically better properties
than prior work.
\name can detect \num{1.5}x overuse flows in one second, whereas prior
approaches fail to detect \num{2}x overuse flows within a timeout of
300 seconds. We demonstrate \name's suitability for high-speed packet
processing with implementations in the DPDK framework and on an FPGA.

\end{abstract}





\section{Introduction}
\label{sec:introduction}
The problem of detecting network flows whose size exceeds a certain share of
link bandwidth, so called \emph{large flows} or \emph{heavy hitters}, has
received significant attention since the seminal paper by Estan and Varghese in
2003~\cite{Estan03} and has recently experienced renewed interest, partly thanks
to the emergence of programmable data-planes \cite{basat2017randomized,
ben2017constant, sivaraman2017heavy, yang2019heavykeeper}. In this formulation
of the heavy-hitter problem, the assumption is that network operators define a
flow size threshold and want to identify all the flows that violate this
threshold.  Previous work~\cite{Estan03, basat2017randomized, ben2017constant,
charikar2002finding, cormode2005improved, Wu14, Wu2018, sivaraman2017heavy,
yang2019heavykeeper} focused on detecting heavy hitters, e.g., flows that are
\num{10}x larger than the threshold such that a large measurement error can be
tolerated. In contrast, the problem of detecting moderately large flows, i.e.,
flows that send slightly more (e.g., \num{1.5}x) than the threshold flow size,
is still in need of an effective solution. Throughout this work, we refer to
these large-flow types as \emph{high-rate} and \emph{low-rate overuse flows},
respectively.

While previous approaches to probabilistic flow-size estimation could be made
arbitrarily accurate if abundant memory was available, these approaches suffer
from a lack of accuracy under stringent constraints regarding memory and
computation. These constraints introduce a large measurement error such that
the detection of overuse flows becomes unreliable. In practice, only flows that
exceed the target threshold by more than the measurement error (which itself can
amount to a significant share of link bandwidth) can be reliably detected,
resulting in many undetected overuse flows.

This failure of probabilistic flow-size monitoring is especially severe under
tight hardware constraints, e.g., on routers that have an aggregate capacity of
several terabits per second (\si{\tbps}), handle millions of concurrent flows,
and thus require high-speed packet processing (on the order of \SI{100}{\ns}
processing time per packet). These heavy constraints only allow for restricted
flow monitoring with an especially large estimation error, which makes the
detection of low-rate overuse flows extremely difficult. At the same time, many
applications that strengthen network
dependability rely on accurate flow-size estimation: Network operators make heavy
use of tools such as network monitoring, traffic engineering (e.g.,
flow-size-aware routing), anomaly detection, Quality of Service (QoS), network
provisioning, and security applications, all of which require accurate insight
into the flow-size distribution.

As one example of a dependability-enhancing application relying on accurate flow-size monitoring,
consider bandwidth-reservation
systems~\cite{sibra2016,rao1998qos,almesberger1997scalable}. These approaches
consist of allocating the available bandwidth on a link to flows according to
purchasable reservations. In case of scarce link capacity (e.g., in case of a
distributed denial-of-service (DDoS) attack), flows with a reservation can
continue sending to the extent of the reserved allowance, whereas other traffic
might be dropped. If using probabilistic flow-size monitoring for allowance
policing, systems that can only detect high-rate overuse flows, but disregard
the detection of low-rate overuse flows, allow an adversary to ``fly under the
radar''. Similar to attacks such as Coremelt~\cite{coremelt} and
Crossfire~\cite{crossfire}, an attacker could wrongfully consume link bandwidth
by creating a large number of flows that only slightly exceed the corresponding
reservation. Thus, detecting low-rate overuse flows is essential to uphold the
guarantees of bandwidth-reservation systems.


The biggest challenge in creating highly accurate flow-size measurement on
high-speed routers is the scarcity of fast memory compared to the enormous
number of flows handled by these routers. Individual-flow resource accounting is
either too expensive (due to the high cost of fast SRAM memory for caches) or
too slow (e.g., keeping per-flow information in
DRAM~\cite{skylake-characteristics, DRAM-latency-2016}). To reduce fast-memory
usage, a line of previous research devises \emph{sketches}, which use a small
number of counters and map every flow to a random subset of these counters. The
size of each flow is then estimated based on the values of the counters
corresponding to that flow~\cite{Estan03, cormode2005improved,
Wu2018,ramabhadran2003efficient,Lall2009}. Flows with an estimated size
exceeding a pre-defined threshold are considered overusing.

Our key insight is that, due to the \emph{high variance of the counter values}
(referred to as \emph{counter noise} in the remainder of this paper), these
shared-counter approaches fail to distinguish low-rate overuse flows from
non-overusing flows. This counter noise originates from two different sources:
(1)~the uneven size of flows within a counter, which leads to non-overuse flows
being mistaken as overuse flows if they are mapped to the same counter as an
overuse flow, and (2)~the uneven number of flows across counters, which leads to
non-overuse flows being mistaken as overuse flows if they are mapped to the same
counter as many other non-overuse flows. Due to counter noise, a sketch cannot
distinguish a \num{1.5}x overuse flow from a non-overuse flow within reasonable
memory limits (cf. \S\ref{sssec:amf}). Hence, the core research challenge
becomes how to effectively counteract the noise while using limited computing
and storage resources.

In this paper, we propose \name, a lightweight detector that can detect low-rate
overuse flows significantly more quickly and reliably than prior approaches,
while conforming to strict requirements regarding time and memory complexity.
\name reduces the counter noise by using a multi-stage approach that is aware of
both the traffic volume and the number of flows in any counter and aggregates
these values over time. Moreover, \name requires fewer operations per packet
than conventional schemes and thereby enables high-speed packet processing,
which we demonstrate with implementations for the DPDK framework~\cite{dpdk} and
for a Xilinx Virtex UltraScale+ FPGA on the Netcope NFB 200G2QL
platform~\cite{netcope-fpga}.

Our evaluation based on both real and synthetic traffic traces shows that \name
outperforms prior work. \name is at least \num{300} times faster than prior
approaches in detecting \numrange{1.5}{2}x overuse flows: \name can reliably
detect \num{1.5}x overuse flows in one second, whereas prior approaches fail
to detect even \num{2}x overuse flows within a time of \SI{300}{\second}.

\section{Problem Definition and Background}
\label{sec:definition}


A \textit{network flow} (or \textit{flow} for short) is a sequence of packets
with common characteristics. For example, NetFlow uniquely defines a flow by
source and destination IP addresses, source and destination transport-layer
ports, protocol, ingress interface and type of service~\cite{Netflow}.

An \textit{overuse flow} is a flow that consumes more bandwidth than it was
permitted by the traversed network. More concretely, if a flow's permitted rate
is $\gamma$ and permitted burst size is $\beta$, then a flow overuses its
allocation if it sends more than $\gamma t + \beta$ in any time interval~$t$. An
$\ell$-fold overuse flow is a flow that sends at rate $\ell\gamma$. The
permitted amount (i.e., the overuse-flow threshold) can be pre-determined by
bandwidth allocation or determined by the router based on its resource usage.

In this section, we review the flow-policing model, with a focus on how the
overuse-flow detection component interacts with other components. Then we
highlight important properties desirable for overuse-flow detectors.


\vspace*{-2pt}
\subsection{Flow Policing Model}
\label{sec:definition:flowpolicingmodel}
\begin{figure}[t]
    \centering
    \includegraphics[width=80mm,clip=true]{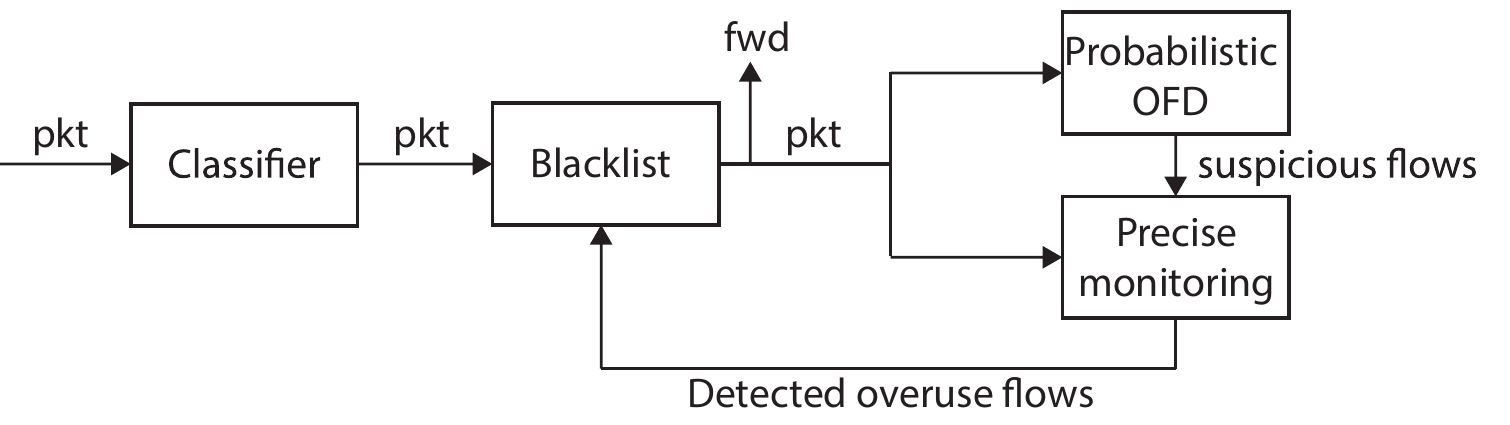}
    \vspace*{-8pt}
    \caption{Router packet processing.
    }
    \label{fig:router-process}
    \vspace*{-12pt}
\end{figure}

A typical flow-policing mechanism consists of the following four components
(shown in Figure~\ref{fig:router-process}): (1)~a flow \emph{Classifier} that
extracts each flow's ID and determines its permitted bandwidth, (2)~a
\emph{Blacklist} that filters out blacklisted flows, (3)~a \emph{Probabilistic
Overuse Flow Detector} (OFD) tasked with finding suspicious flows that could
potentially be overusing, and (4)~a \emph{Precise monitoring} component which
analyzes individual suspicious flows to determine which ones are actually
misbehaving and should be added to the blacklist. The precise-monitoring
component can access a limited amount of fast memory (e.g., on the order of the
amount used in the overuse flow detector). This limited fast memory restricts
the number of suspicious flows that can be simultaneously monitored by the
detector, leading to false negatives.

While stateful monitoring of individual flows is practical at the edge of the
network, stateful monitoring has an untenable fast-memory consumption on routers
with Tbps capacity~\footnote{
As the CAIDA dataset suggests flow concurrency of 10 million flows on a switch
supporting an aggregate bandwidth of 1 Tbps\cite{caida201810}, individual-flow
monitoring can be expected to require 80MB of fast memory, assuming 4 byte flow
IDs and 4 byte counters.}.

Moreover, schemes with per-flow state in fast memory are vulnerable to memory
exhaustion attacks in which an attacker creates a high number of flows and
thereby depletes the available fast memory. Hence, only probabilistic monitoring
is a viable option on high-speed routers.




\vspace*{-2pt}
\subsection{Desired Properties}
To ensure accurate and timely detection of overuse flows without affecting the
regular packet processing, an overuse-flow detection algorithm should satisfy
the following properties:

\paragraph{High-speed packet processing on routers}
High-speed packet processing is required for a system that is deployed in the
fast path of Internet routers (i.e., packet processing). However, high-speed
memory that enables the necessary packet-processing frequency (e.g., SRAM) is
expensive. Therefore, even core switches provide only a very limited amount of
fast memory to ensure sustainable hardware cost.
DRAM, albeit cheaper, is too slow to be accessed for every packet, but may be
used for router slow-path operations (i.e., back-end processing on a CPU)
without impacting fast-path forwarding. Moreover, even when accessing only SRAM
in the router fast path, memory accesses should be employed economically to keep
up with line rate.

\paragraph{Low false-positive and low false-negative rates}
In the context of overuse flow detection, a \emph{false fositive} (FP) is a
misclassification of a non-overuse flow as an overuse flow. Conversely, a
\emph{false negative} (FN) is a misclassification of an overuse flow as a
non-overuse flow, i.e., the failure of detecting an overuse flow. Although the
precise-monitoring component of a flow-policing mechanism can prevent
non-overuse flows from being falsely blacklisted, the detector itself should
still ensure a low FP rate to not evict suspicious flows in
the precise-monitoring component, thus increasing FN rates.

\paragraph{Detection of low-rate overuse flows}
Overuse flows that are sending slightly above their permitted sending rate
should be detected with high probability after a short amount of time. Our goal
is to detect flows within at most seconds that are sending at \num{1.5} their
permitted rate; current state-of-the art algorithms typically assume
\numrange{10}{1000} fold sending rates of overuse flows when budgeting
resources.

\subsection{Overuse Flows vs. Large Flows}
\label{ssec:definition:comparison}
This work aims to efficiently detect overuse flows, and is inspired by
algorithms for detecting \emph{large flows}, i.e., flows which use a significant
fraction of the link bandwidth.

However, it is important to note the significant differences between the two
problems. In our context, large flows correspond to high-rate overuse flows,
i.e., flows sending at rates at least \numrange{10}{1000} times higher than the
average flow (for example, flows violating TCP fairness). The goal of previous
work is to quickly identify the large flows to throttle or block them, thus
preventing them from harming the other flows. These algorithms have a (more or
less explicit) threshold above which a flow is considered large, but below which
flows are allowed to send. This threshold is usually up to three orders of
magnitude larger than the average flow's sending rate---so a few flows sending
around the threshold rate would rapidly exhaust link capacity and lead to
congestion.




In the rest of this section, we briefly introduce two kinds of large-flow
detection algorithms, namely sketch-based approaches and selective
individual-flow monitoring. In the following, we discuss why they are inadequate
for solving overuse-flow detection in our target scenario.


\subsubsection{Sketches}
\label{sssec:amf}
Individual-flow monitoring tracks the size of flows with a \emph{counter} per
flow, i.e., a memory cell that is increased by the packet size every time a
packet of the corresponding flow arrives. To monitor flows using limited fast
memory, \emph{sketches} use each counter to track multiple flows. Two algorithms
in this category are the \emph{Count-Min (CM) Sketch} \cite{cormode2005improved}
(also known as Multistage Filters~\cite{Estan03}) and \emph{Adaptive Multistage
Filters} (AMF)~\cite{Estan2003a}. They rely on a relatively simple concept,
which we illustrate at the example of the CM Sketch.

In the CM Sketch, flows are randomly mapped to counters, and each counter
aggregates the volume of all flows that are assigned to it. If a large flow is
mapped to a certain counter, this counter value is expected to be higher than
the other counters as it includes the contribution of the large flow. To
increase precision, the CM Sketch uses multiple \emph{stages}, i.e., multiple
counter arrays, and for each stage the flows are mapped to counters in a
different way (e.g., using different hash functions). The CM Sketch classifies a
flow as large if and only if the minimum value of counters to which the flow is
mapped exceeds a certain threshold. For non-large flows, the probability that
all associated counters exceed the threshold is low, decreasing exponentially in
the number of stages.

However, achieving high accuracy with a CM Sketch is only possible with an
untenable amount of memory. According to the theoretical work on the CM
Sketch~\cite{cormode2005improved}, the measurement error of the CM Sketch can be
related to the amount of available memory. Concretely, a CM Sketch with $\lceil
\ln(1/\delta) \rceil$ stages, each with~$\lceil e/\epsilon \rceil$ counters,
guarantees a probability of less than $\delta$ that the overestimation error
amounts to more than a share~$\epsilon$ of total traffic. Assuming that a switch
with an aggregate bandwidth of \SI{1}{\tbps} handles \num{10} millon
flows~\cite{caida201810} and that the overestimation should almost never exceed
\num{50}\% of the average-flow size (hence, $\delta = 0.01$ and $\epsilon =
0.5/10^{-7}$), then the CM Sketch would require around \num{250} million
counters. This memory consumption is even higher than allocating a counter per
flow, which demonstrates that the CM Sketch is inaccurate on high-capacity
routers.

Moreover, conventional sketches have been shown to be too inefficient to keep up
with usual line speeds~\cite{liu2019nitrosketch}. In order to achieve line rate,
accuracy has to be traded for speed, which exacerbates the estimation error of
sketches. In this work, we attempt to refine the sketch-based approach in order
to achieve high accuracy with low processing complexity.

\subsubsection{Selective Individual-Flow Monitoring}
\label{sssec:eardet}
Another category of large-flow detection algorithms dynamically selects a subset
of flows for individual-flow monitoring. In the following section, we will
illustrate the general idea of these schemes using the example of
EARDet~\cite{Wu14}, which is one specimen of this category. Other examples
include HashPipe~\cite{sivaraman2017heavy} and HeavyKeeper
\cite{yang2019heavykeeper}.

EARDet is based on the Misra--Gries (MG) algorithm~\cite{Misra1982}, which finds
the exact set of frequent items (i.e., items making up more than a $1/k$-share
of the stream) in two passes with limited counters. At a high level, the MG
algorithm uses an array of counters to track frequent item candidates. By
adjusting the counter values and associated items, the MG algorithm guarantees
that every frequent item will occupy one counter after the first pass. The
second pass is nevertheless required to remove falsely included infrequent
items.

For each item in the stream, the MG algorithm adjusts the counters as follows.
It first checks whether the item has occupied a counter in the array. If so, the
corresponding counter will be increased by one. Else, if there is a non-occupied
counter, the MG algorithm will assign that non-occupied counter to track this
item (and also increase it by one). Otherwise, it will decrease all counters by
one. The intuition is that infrequent items, should they be assigned to a
counter, are likely to be evicted (i.e., their counter becomes zero) very
quickly. By contrast, frequent items (which are already more likely to be
assigned to a counter to begin with) are guaranteed to remain assigned to that
counter, since their frequency compensates the occasional counter decreases.

EARDet enhances the MG algorithm to identify large network flows in \emph{one
pass}. EARDet is based on two flow specifications of the form $\gamma t +
\beta$, where $\gamma$ is the allowed rate and $\beta$ is the allowed burst
size. The adaptations guarantee that a flow sending less traffic than $\gamma_l
t + \beta_l$ during any time window with length $t$ will not be falsely blocked
(no false positive), and all large flows sending more than $\gamma_h t +
\beta_h$ in any time window with duration $t$ will be caught (no false
negative).


To catch overuse flows, EARDet could be configured to have $\gamma_h t +
\beta_h$ set to the permitted bandwidth. However, given a low permitted
bandwidth, this approach either demands many counters or suffers from high false
negatives, as EARDet recommends using at least $\frac{\mathit{link
BW}}{\gamma_h} - 1$ counters to achieve guaranteed detection. If $\gamma_h$
equals the maximum size of a non-overuse flow, fast memory would need to
accommodate almost a counter per flow, which is infeasible on routers that
handle millions of flows (cf.\ \S\ref{sec:definition:flowpolicingmodel}).

In addition to their high fast-memory consumption, the overhead of selective
individual-flow monitoring, i.e., continuously deciding which flows to monitor
closely, is prohibitively expensive in terms of processing complexity, which
results in an insufficient throughput of such schemes (cf.
\S\ref{sec:scalability}).

\section{\name Algorithm}
\label{sec:algorithm}
\newcommand{\vcnt}{\overline{\accum}}
\newcommand{\vsiz}{\overline{\counter}}

\name is a novel design approach for a probabilistic overuse flow detector
(OFD), the core component of a flow policing model in router packet processing.
We first provide an overview of \name (\S\ref{ssec:alg:overview}) before
describing the design in more detail (\S\ref{ssec:update}--\ref{ssec:sampler}).
We also provide a complexity analysis of the algorithm
(\S\ref{ssec:complexity}).

\begin{table}[t]
    \centering
    \caption{Notation used in this paper.}
    \vspace*{3pt}
    \begin{tabular}{cl}
        \toprule
        Symbol                & Description                                 \\
        \midrule
        $\flownum$            & Number of flows                             \\
        $\fsgamma$, $\fsbeta$ & Rate and burst threshold flow specification \\
        $\ratio$              & Overuse ratio                               \\
        $\parts$              & Number of minor cycles since last reset     \\
        $\fastfreq$           & Number of minor cycles per second           \\
        $\slowfreq$           & Number of minor cycles per major cycle      \\
        $\resetcycle$         & Reset cycle                                 \\
        $\lambda$             & Sample rate                                 \\
        $\fwidth$             & Number of counters in fast memory                 \\
        $\fmwidth$            & Number of precisely monitored flows         \\
        $\estim$              & Flow-size estimate                          \\
        $\accum$              & Accumulated flow size                       \\
        $\counter$            & Accumulated flow count                      \\
        \bottomrule

    \end{tabular}
	\label{tbl:notation}
\end{table}

\subsection{Overview}
\label{ssec:alg:overview}

\begin{figure}[t]
    \centering
    \includegraphics[width=0.8\columnwidth,clip=true]{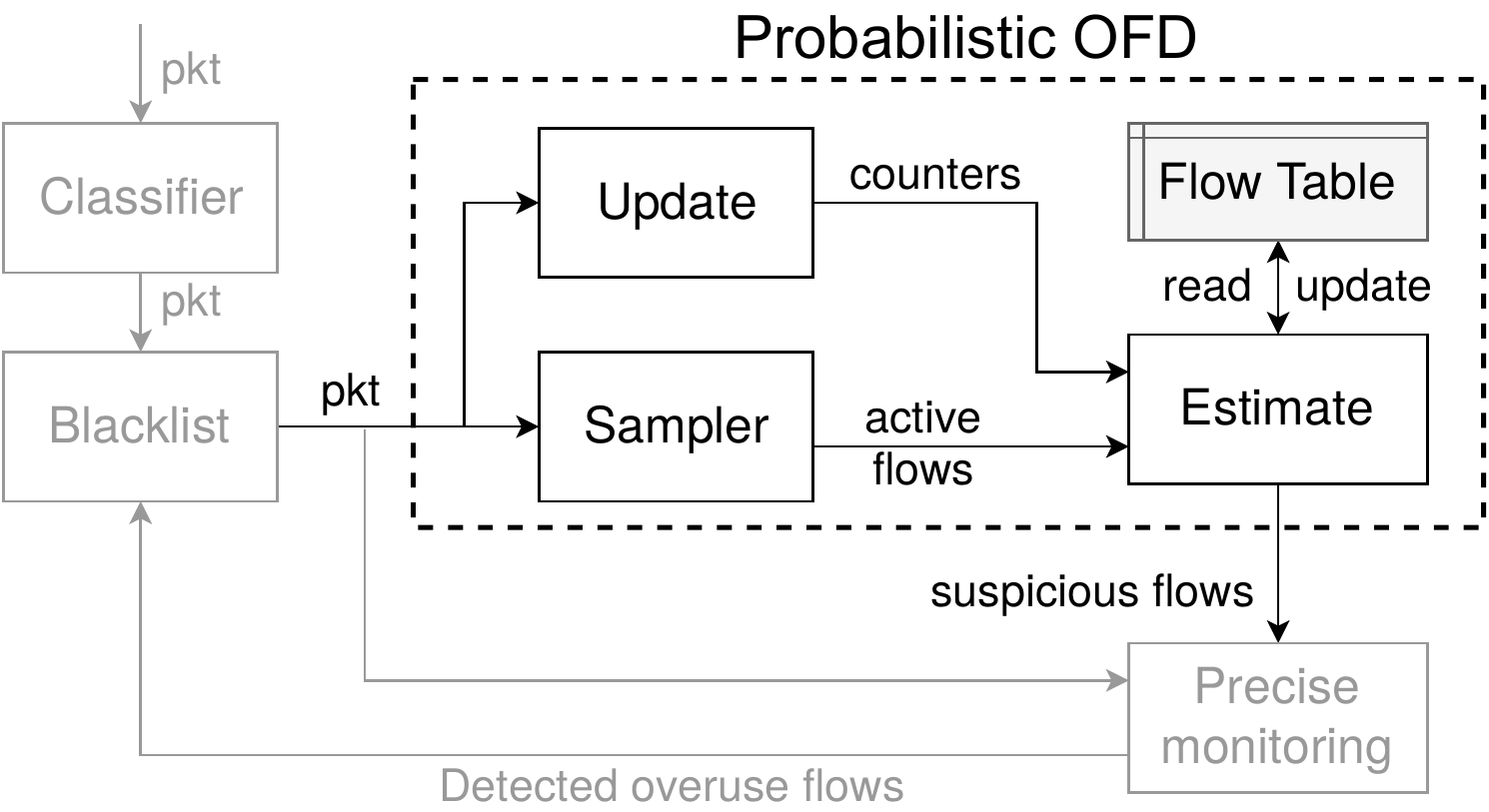}
    \vspace*{-5pt}
    \caption{Details of the structure and information flow of the \name
        Probabilistic OFD component (cf. also Figure~\ref{fig:router-process}).}
    \label{fig:architecture}
    \vspace*{-8pt}
\end{figure}

In this section, we describe our design for the probabilistic OFD component
depicted in Figure~\ref{fig:router-process}. As Figure~\ref{fig:architecture}
shows, \name OFD contains four components: the \emph{update} algorithm, the
\emph{estimate} algorithm, the \emph{sampler}, and the \emph{flow table}.
Packets not rejected by the blacklist are forwarded to the \emph{update}
algorithm and the \emph{sampler}. The \emph{estimate} algorithm consumes their
output, updates the \emph{flow table} and creates a list of suspicious flows for
precise monitoring.

\begin{figure*}[tb]
    \centering
    \includegraphics[width=0.8\linewidth,clip=true]{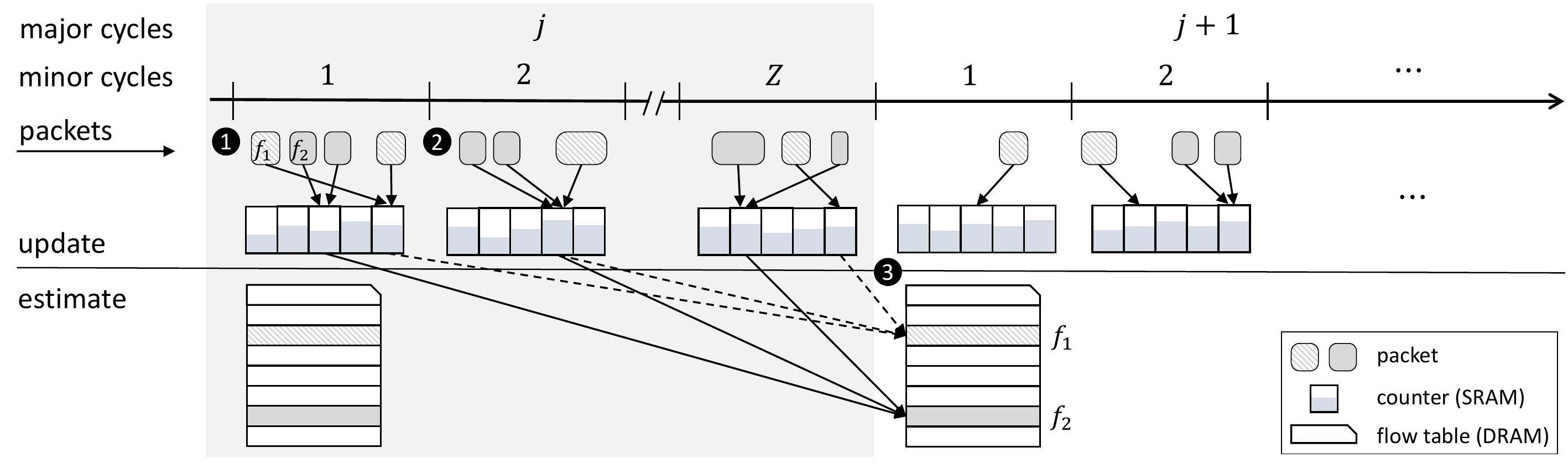}
    \caption{\name Timeline. \ding{182} A hash value is computed on the flow ID
        of the incoming packet, and the corresponding counter is increased by
        the packet size. \ding{183} A different set of counters and hash function
        are used when the minor cycle changes. \ding{184} After $\slowfreq$ minor
        cycles, the \emph{estimate} algorithm aggregates the counter values and
        tries to identify a group of overuse flows.}
    \label{fig:timeline}
\end{figure*}

The \name \emph{update} algorithm targets one short time interval (e.g.,
\SI{12.5}{\ms}) at a time, which we call \emph{minor cycles}. For each
minor cycle, the \emph{update} algorithm collects aggregated traffic information
over groups of flows by using a single counter array. For every packet, \name
maps the packet flow ID to one counter in the current counter array and
increases that counter by the packet size. At the end of the minor cycle, the
counter array is passed to the \emph{estimate} algorithm.

The \emph{estimate} algorithm operates on a larger time scale, at intervals
(e.g., with duration \SI{250}{\ms}) that we call \emph{major cycles} (see
Figure~\ref{fig:timeline}): every time a major cycle is concluded, the
\emph{estimate} algorithm analyzes the counter arrays stored from all minor
cycles during the major cycle, extracts estimates for the bandwidth utilization
of every flow, and stores the estimates in a \emph{flow table}. From this, the
algorithm produces a list of suspected overuse flows, which is handed to the
precise-monitoring component. The \emph{sampler} provides a list of active flows
which the \emph{estimate} algorithm uses in its analysis.

In summary, the \emph{estimate} algorithm uses the the sequence of counter
arrays generated by the \emph{update} algorithm to create a final flow estimate
in each major cycle. Figure~\ref{fig:timeline} visualizes how these algorithm
components interact. By depicting packets of two different flows, the figure
shows how flows are mapped to different counters in each minor cycle. Based on a
packet's flow ID, the \emph{update} algorithm increases the associated counter
by the packet size. Every major cycle, the \emph{estimate} algorithm first
updates the \emph{flow table} based on the counter arrays generated in the most
recent $\slowfreq$ minor cycles, recomputes the estimate for every active flow,
and creates a \emph{watchlist} from the $\fmwidth$ largest flows. The flows on
the \emph{watchlist} undergo precise monitoring in the subsequent major cycle.
This precise monitoring is performed by the leaky-bucket
algorithm~\cite{turner1986new}, which detects any violation of a flow
specification in form of~$\gamma t + \beta$ (cf. \S\ref{sec:definition}) without
false positives. Any flow that is found misbehaving by precise monitoring can
thus be blocked by inserting it into the \emph{blacklist}.

In the following, we will explain the design of the \emph{update} algorithm
(\S\ref{ssec:update}), the \emph{estimate} algorithm (\S\ref{ssec:estimator}),
and the \emph{sampler} (\S\ref{ssec:sampler}). The pseudo code of the whole
system is presented in Algorithm~\ref{alg:main}. The relevant notation is
presented in Table~\ref{tbl:notation}.

\subsection{Update Algorithm}
\label{ssec:update}

Similar to a sketch, the \name \emph{update} algorithm maps each flow to one
counter in a counter array. Each counter tracks the aggregate bandwidth of all
the flows mapped to that counter during a minor cycle. In each minor cycle, the
association of flows to counters is randomized by changing the hash function for
every minor cycle. When a minor cycle ends, the counter array is moved to main
memory and an empty counter array is initialized in fast memory.

More formally, let $\mathit{ctr}_{j, k}$ be the counter array generated in the
$k$-th minor cycle of major cycle $j$, and $H_{j,k}$ be the hash function used
in that minor cycle. The value of $\mathit{ctr}_{j, k}[x]$ is the total packet
sizes of flows that have been mapped to $x$ by~$H_{j,k}$, i.e., all flows $f$
for which $H_{j,k}(f) = x$.

\subsection{Estimate Algorithm}
\label{ssec:estimator}

At the end of a major cycle, which contains a certain number~$Z$ of minor
cycles, the \emph{estimate} algorithm performs a flow-size estimation for every
flow asynchronously (e.g., in user space of the router), while the \emph{update}
algorithm continues to aggregate traffic information. The flow-size estimate
builds on two values: the \emph{volume sum} and the
\emph{cardinality sum}.

For the volume sum, the \emph{estimate} algorithm sums up the values of all the
counters to which a flow was mapped. This aggregation over time reduces the
counter noise in the sense of uneven flow size within the same counter:
intuitively, an overuse flow, unlike a non-overuse flow, will be consistently
associated with large counter values, which results in a large volume sum for
that flow.

Although using multiple counters can reduce counter noise, it is neither
sufficient nor innovative, as the Count-Min Sketch \cite{cormode2005improved}
uses the same idea (although by applying different hash functions concurrently
instead of sequentially) and delivers insufficient performance.
Indeed, the key to reducing counter noise lies in the cardinality sum. In order
to compute this sum, an active-flow list is consulted to compute how many flows
are associated with each counter within each minor cycle, i.e., the counter
cardinality. For every flow, the \emph{estimate} algorithm sums up the
cardinality values of all counters associated with the flow. The cardinality sum
reduces the distortion created by the varying cardinalities of counters:
intuitively, an overuse flow will be associated with a large counter value even
when the number of flows in that counter is small.

When dividing the volume sum by the cardinality sum, the strongest increases in
a flow's estimate are produced when the flow is mapped to high-value counters
that contain a small number of flows. Indeed, flows with these characteristics
are highly likely to be the largest flows among the investigated flows and are
thus candidates for more precise monitoring.

Formally, after major cycle $j$, we define the estimate of a flow $f$ to be
$\overline{U}_f^{(j)} = \accum_f^{(j)} / \counter_f^{(j)}$, where
\begin{subequations}
    \setlength{\abovedisplayskip}{0pt}
    \setlength{\belowdisplayskip}{0pt}
    \begin{align}
        \accum_f^{(j)}   & = \sum_{j' \in J_f^{(j)}} \sum_{k=1}^\slowfreq \mathit{ctr}_{j',k} [H_{j',k}(f)]             \\
        \counter_f^{(j)} & = \sum_{j' \in J_f^{(j)}} \sum_{k=1}^\slowfreq \left|\mathit{ctr}_{j',k}[H_{j',k}(f)]\right|
    \end{align}
\end{subequations}
where $J_f^{(j)}$ contains all major cycles $j' \leq j$ in which flow $f$ was active. The term $|\mathit{ctr}_{j', k}[x]|$
denotes the number of flows that have been mapped to counter $x$ in minor cycle $k$ of major cycle $j'$
(counter cardinality). $\accum_f^{(j)}$ is the value aggregate of the counters that $f$ has
been mapped to (volume sum) and $\counter_f^{(j)}$ is the summed count of the flows
in these counters (cardinality sum).
In order to avoid preserving counter arrays from
past major cycles, the terms $\accum_f$ and
$\counter_f$ are kept in the flow table
and updated after every major cycle, i.e., \begin{equation}
    \mathit{table}[f].\accum \leftarrow \mathit{table}[f].\accum +
    \sum_{k=1}^\slowfreq ctr_{j,k} [H_{j,k}(f)]
\end{equation} and analogously for $\counter_f^{(j)}$.
These updates are made for all flows~$f$ that were active in the most recent major cycle and are thus
in the active-flow list generated by the sampler (cf.~\S\ref{ssec:sampler}).





These estimates need to be adjusted when some flows send intermittently. For
example, suppose flow $f_1$ sends $x$ \si{\GB} in the first and the third major
cycle and nothing in the second major cycle, and flow $f_2$ sends $x$ \si{\GB}
from the first to the third major cycle, i.e., $J_1^{(3)} = \{1, 3\}$ and
$J_2^{(3)} = \{1, 2, 3\}$. Then $\accum_1^{(3)}/\counter_1^{(3)}$ and
$\accum_2^{(3)}/\counter_2^{(3)}$ will be almost the same. Suppose all counters
contain exactly $y$ flows, then $\accum_1^{(3)}/\counter_1^{(3)} =
\frac{x+x}{y+y} = \frac{x}{y} = \frac{x+x+x}{y+y+y} =
\accum_2^{(3)}/\counter_2^{(3)}$. However, the total traffic sent by flow $f_2$
in these three cycles is actually 1.5 times larger than flow $f_1$ and should
result in a higher flow-size estimate.

To fix this problem, we reduce a flow-size estimate $\overline{U}_f$ relative to
the number of major cycles where flow~$f$ was not active, i.e.,
\begin{equation}
U_f^{(j)} = \frac{|J_f^{(j)}|}{j} \cdot \frac{A_f^{(j)}}{C_f^{(j)}}
\end{equation}
To enable this computation, the flow table must track~$|J_f|$ for every
flow~$f$.

Another issue is that as  $\accum_f$ and $\counter_f$ are accumulated, the
\emph{estimate} algorithm is actually computing their average over time.
An attacker can take advantage of this approach by sending a low-rate traffic in
the beginning for a period of time, and then start sending bursty traffic. It
may not be detected by our system as its long-term average looks the same as a
non-overuse flow, so old values must be discarded at some point. Therefore, we
define the reset cycle $\resetcycle$, and clear all the data every $\resetcycle$
minor cycles.
This reset may sound risky, as an overuse flow could send its traffic around
the reset point so that its estimated size is reset before being detected by our
system. However, an attacker does not know the reset point. Moreover, even if an
attacker could infer the reset point, the overuse traffic sent by a flow with
such a strategy is bounded, as we show in the mathematical analysis (\cref{sec:analysis}).

\subsection{Sampler}
\label{ssec:sampler}


Clearly, \name requires a list of active flows for which an estimate must be
computed.
In order to generate such an active-flow list, we use sampling and limit the number of sampled packets per second to be $\lambda$.
\name considers a randomized sampling period, which is a random variable of an
exponential distribution with mean $\frac{1}{\lambda}$. This randomization
prohibits an attacker flow from circumventing the sampling by sending at the
appropriate moments.

Having an active-flow list for a major cycle~$j$ also allows to compute the
cardinality $|\mathit{ctr}_{j,k}[x]|$ of counters in the \emph{estimate}
algorithm, namely by counting how many active flows were mapped to each counter
with the respective hash function~$H_{j,k}$ for any minor cycle~$k$. An
alternative to this reconstruction of counter cardinality would consist in
measuring counter cardinality within the \emph{update} algorithm, for example
using a Bloom filter~\cite{bloom1970space} or HyperLogLog
register~\cite{flajolet2007hyperloglog} per counter. However, as fast memory is
the bottleneck resource, additional computational complexity in the
\emph{estimate} algorithm is preferable to estimating cardinality in fast
memory.



\begin{algorithm}[t!]
    \caption{\name algorithm.}
    \label{alg:main}
    \small
    \begin{algorithmic}[1]
        \Procedure{Process}{$pkt$}
        \If{$pkt.flowID \in blacklist$}
        \State \Return
        \EndIf
        \If{$pkt.flowID \in watchlist$}
        \State \Call{Monitor}{$pkt$}
        \EndIf
        \State $j \gets$ \Call{GetMajorCycle}{\null}
        \State $k \gets$ \Call{GetMinorCycle}{\null}
        \State \Call{Sampler}{$pkt$}
        \State \Call{Update}{$pkt$, $j$, $k$}
        \If{$\slowfreq \cdot j \geq \resetcycle$}
        \State \Call{Reset}{\null}
        \EndIf
        \EndProcedure
        \Procedure{Sampler}{$pkt$}
        \If{$\mathit{current\_time} \geq \mathit{sample\_time}$}
        \State $activeFlow \gets activeFlow \cup \{ pkt.flowID \}$
        \State $u$ samples uniformly from $\mathcal{U}(0, 1]$
        \State $sample\_time \gets sample\_time - \frac{\ln{u}}{\lambda}$
        \EndIf
        \EndProcedure
        \Procedure{Update}{$pkt$, $j$, $k$}
        \State $x \gets H_{j,k}(pkt.flowID)$
        \State $\mathit{ctr}_{j,k}[x] \gets ctr_{j,k}[x] + pkt.size$
        \EndProcedure
        \Procedure{Estimate}{\null}
        \State $j \gets$ \Call{GetMajorCycle}{\null} - 1
        \For{$k = 1$ \textbf{to} $\slowfreq$}
        \For{$f \in activeFlow$}
        \State $x \gets H_{j,k}(f)$
        \State $numFlow[x] \gets numFlow[x] + 1$
        \EndFor
        \For{$f \in activeFlow$}
        \State $x \gets H_{j,k}(f)$
        \State $A[f] \gets A[f] + ctr_{j,k}[x]$
        \State $C[f] \gets C[f] + numFlow[x]$
        \EndFor
        \EndFor
        \State $\flownum \gets |activeFlow|$
        \For{$f \in activeFlow$}
        \State $table[f].A \gets table[f].A + A[f]$
        \State $table[f].C \gets table[f].C + C[f]$
        \State $table[f].\mathit{numJ} \gets table[f].\mathit{numJ} + 1$
        \EndFor
        \State $activeFlow \gets \emptyset$
        \State \Return $\fmwidth$ flows with largest
        $\frac{table[f].\mathit{numJ}}{j}\cdot\frac{table[f].A}{table[f].C}$
        \EndProcedure
    \end{algorithmic}
\end{algorithm}

\subsection{Complexity Analysis}
\label{ssec:complexity}
In summary, the \emph{update} algorithm uses $O(W)$ fast-memory entries,
where~$W$ is the width of a counter array. The number of read and write
operations is linear in the number of packets. The estimate algorithm uses
$O(\slowfreq\fwidth+\flownum)$ entries in DRAM, and the number of accesses is
$O(\slowfreq\flownum)$. 

\subsubsection{Time complexity}
In the \emph{update} algorithm, each packet requires a single read and a single
write operation to fast memory. Moreover, the algorithm requires a single hash
function computation, which results in the major advantage  that \name achieves
line rate (see \S\ref{sec:scalability}), whereas other sketch-based algorithms
update multiple counter arrays per packet and therefore fall short of that
goal~\cite{liu2019nitrosketch}. The hash computation itself can be performed in
hardware or using an efficient software implementation such as the
\emph{murmur3} hash function.

The \emph{estimate} algorithm uses multiple accesses to main memory. However,
the number of accesses is linear in the number of active flows, which may be
much smaller than the number of packets. In each major cycle, we need to sum up
the corresponding counters in each minor cycle for each flow. Suppose there are
$\flownum$ active flows, then there are $O(\slowfreq\flownum)$ DRAM reads to
compute the updates to the estimate components.
After obtaining these values, we need to update the flow table. For each active
flow, the algorithm performs one lookup and update to the hash table. We use a
Cuckoo hash table~\cite{pagh2001}, which has worst-case constant lookup time.
Although the worst-case insertion time of the Cuckoo table might be long, its
expected complexity is amortized constant. Additionally, the number of
insertions is much lower than the number of lookups as each flow will be
inserted only once when it's first seen. Therefore, the update takes
$O(\flownum)$ time.

In the sampler, for every sampled packet, there is one insertion to the
active-flow list stored in DRAM, for which we again used a Cuckoo hash table. By
properly setting the table capacity and sample rate~$\lambda$, our experiments
show that the sampler is still fast enough to keep up with line speeds.

\subsubsection{Space Complexity}
Only counters of the current minor cycle reside in fast memory, which is
$O(\fwidth)$ and depends on the size of a counter. Our analysis shows that, if
all flows send almost at threshold rate, a small $\fwidth$ (e.g., \num{8192})
can lead to considerable detection delay.

Other counters are kept in main memory before being handled by the
\emph{estimate} algorithm, which takes $O(\slowfreq\fwidth)$ space. The
active-flow list and the flow table are also in DRAM. The active-flow list
requires $O(\flownum)$ entries and the space complexity of the flow table is
also linear to $\flownum$ (using a Cuckoo table). Therefore, the total number of
main memory entries used (including counters) is $O(\slowfreq\fwidth+\flownum)$.

\section{Evaluation}

\label{sec:evaluation}
The evaluation of \name is conducted through two implementations and a
simulation. First, the behavior of \name in a real-world environment is
evaluated using the implementations and a testbed that supports up to
4x\SI{40}{\gbps} traffic volume. Second, to evaluate the accuracy of \name and
compare it to EARDet, AMF, HeavyKeeper and HashPipe, simulations with a traffic
volume equivalent to 4x\SI{100}{\gbps} are used.

\subsection{Implementation}
For the scalability experiments with DPDK, we implemented \name in~C on the
Intel DPDK framework~\cite{dpdk}. The application uses $n$ worker threads that
execute the \fpart and a separate thread running the \spart every major cycle.
The major and minor cycle indices are computed based on a monotonic clock with
nanosecond resolution.

For further scalability experiments, we also implemented the \fpart of \name on
a Xilinx Virtex UltraScale+ FPGA on the Netcope NFB 200G2QL
platform~\cite{netcope-fpga} containing two \SI{100}{\gbps} NICs. Presenting the
detailed contributions of the FPGA implementation would exceed the scope of this paper.
Therefore, a separate paper illustrating the design of the FPGA implementation
has been submitted to a specialized conference~\cite{anonymous2020fpga}.

In order to test the accuracy of \name, we evaluated \name and other algithms in
simulations using Rust (\name, EARDet, AMF) and Golang (HeavyKeeper, HashPipe).

\subsection{Experiment Setup}
\paragraph{Test Setup}
The testbed for the DPDK scalability experiment consists of two machines
connected using 4x\SI{40}{\gbps} Ethernet connections. The traffic is generated
using a dedicated traffic generator (Spirent TestCenter N4U~\cite{spirent}) and
sent to a commodity machine with an Intel Xeon E5-2680 CPU. For the simulations,
we execute \name on an Intel Xeon 8124M machine and process synthesized traffic
equivalent to 4x\SI{100}{\gbps}.

\paragraph{Traffic Generation}
To evaluate the scalability of \name with respect to flow volumes, we generated
network traces with uniform packet sizes and based on an iMix traffic
distribution (avg. size: \SI{353}{\B})~\cite{RFC6985}.





\subsection{Evaluation Metric: Detection Delay}
The \emph{detection delay} of an overuse flow is the time elapsed between the
first violation of the flow specification and the time the flow is caught by a
detector. A detection delay longer than the simulation timeout results in a
\emph{false negative}.

When presenting detection rates and false negatives, it is always crucial to
present the corresponding false positive rate, i.e., the number of non-overuse
flows incorrectly marked as malicious. We note, however,  that \name is designed
to have no false positives, because benign flows that happen to be flagged as
suspicious by the \spart will be exonerated by precise monitoring.

\subsection{Parameter Selection}
Several of \name's parameters can be tuned to fit the hardware restrictions of a
specific deployment. In the following, we describe how to experimentally
determine these parameters so that they comply with the hardware's computational
constraints even under the worst-case traffic patterns.

To experimentally determine the sampling rate, we increase the sampling rate
of the sampler until its CPU core is fully utilized, such that the accuracy of
the flow ID list is maximized. With this determined sampling rate and an
estimated maximum number of flows, we calculate the maximum number of major
cycles per second such that there are enough samples between two executions of
the \spart to build the flow ID list with the desired accuracy. Finally, we
increase the number of minor cycles per second until the CPU core of the \spart
is fully utilized. Table~\ref{default-param} summarizes these hardware-related
parameters used in our experiments.

The flow monitor is set to monitor 64 flows simultaneously in all experiments,
which is small and fast enough to keep up with a high-bandwidth link. Finally,
with the determined parameters in Table \ref{default-param}, the reset cycle
parameter $\resetcycle$ in each experiment is adjusted to achieve \num{95}\%
detection probability (detailed in the mathematical analysis in~\cref{sec:analysis})
under the given experiment setting.

\begin{table}[h]
    \centering
    \caption{Hardware-related \name parameters.
    }\label{default-param}
    \begin{tabular}{lr}
        \toprule
        Parameter                 & Value             \\
        \midrule
        Sampling rate ($\lambda$) & $2.1\cdot10^6$ samp./s \\
        Number of minor cycles    & \num{64} cycles/s       \\
        Number of major cycles    & \numrange{1}{4} cycles/s     \\
        \bottomrule
    \end{tabular}
    \vspace*{-10pt}
\end{table}

\subsection{Comparison: Fully Utilized Traffic Trace}
We first evaluate \name in a setting where every flow sends at a rate close to
the maximum allowed threshold, fully utilizing the reserved bandwidth.

We simulate a configuration with 4x\SI{100}{\gbps} links with an aggregate
number of \num{130000} flows, where each flow requires \SI{3}{\mbps}, e.g., for
high-quality video streaming. Then, a misbehaving flow with an overuse ratio
$\ratio$ is injected. Each detector allocates \num{16448} counters in fast
memory. \name, AMF, HashPipe and HeavyKeeper use \num{64} counters (out of
\num{16448}) as flow monitors. To optimize detector performance for AMF,
HashPipe and HeavyKeeper, the fast-memory counters are structured as counter
arrays. \name is configured to run \num{4} cycles of the \spart per second,
which reaches the computation limit on our machine.

\begin{figure}[t]
    \setlength\abovecaptionskip{-0.1\baselineskip}
    \centering
    \vspace*{-3pt}
    \subfigure[$N = 130K$] {
    	\includegraphics[trim=0 0 0 0,clip,width=0.85\columnwidth]{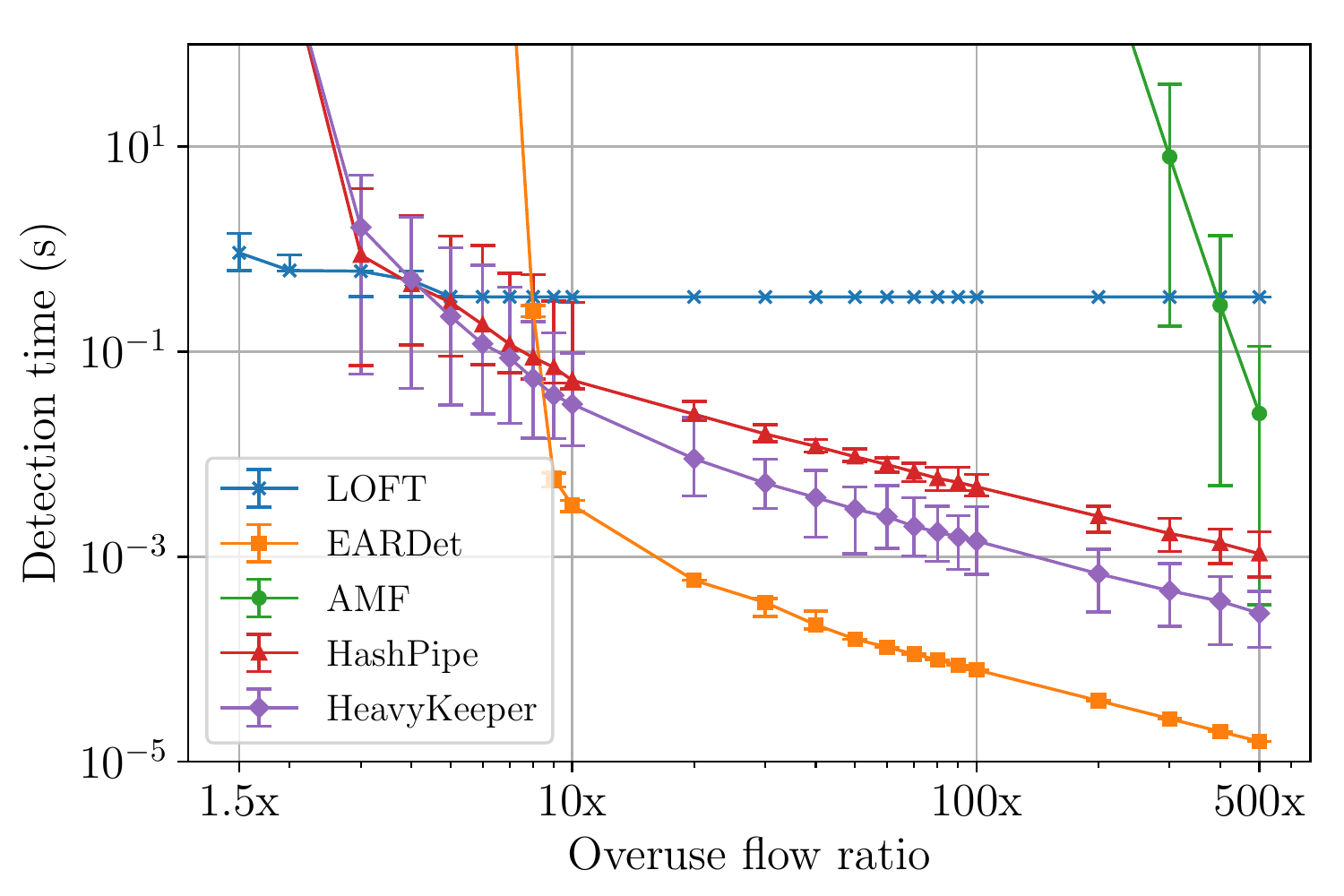}
    	\label{fig:graph_cmp}
	} \\[-5pt]
	\subfigure[$N = 10M$] {
		\includegraphics[trim=0 0 0 10,clip,width=0.85\columnwidth]{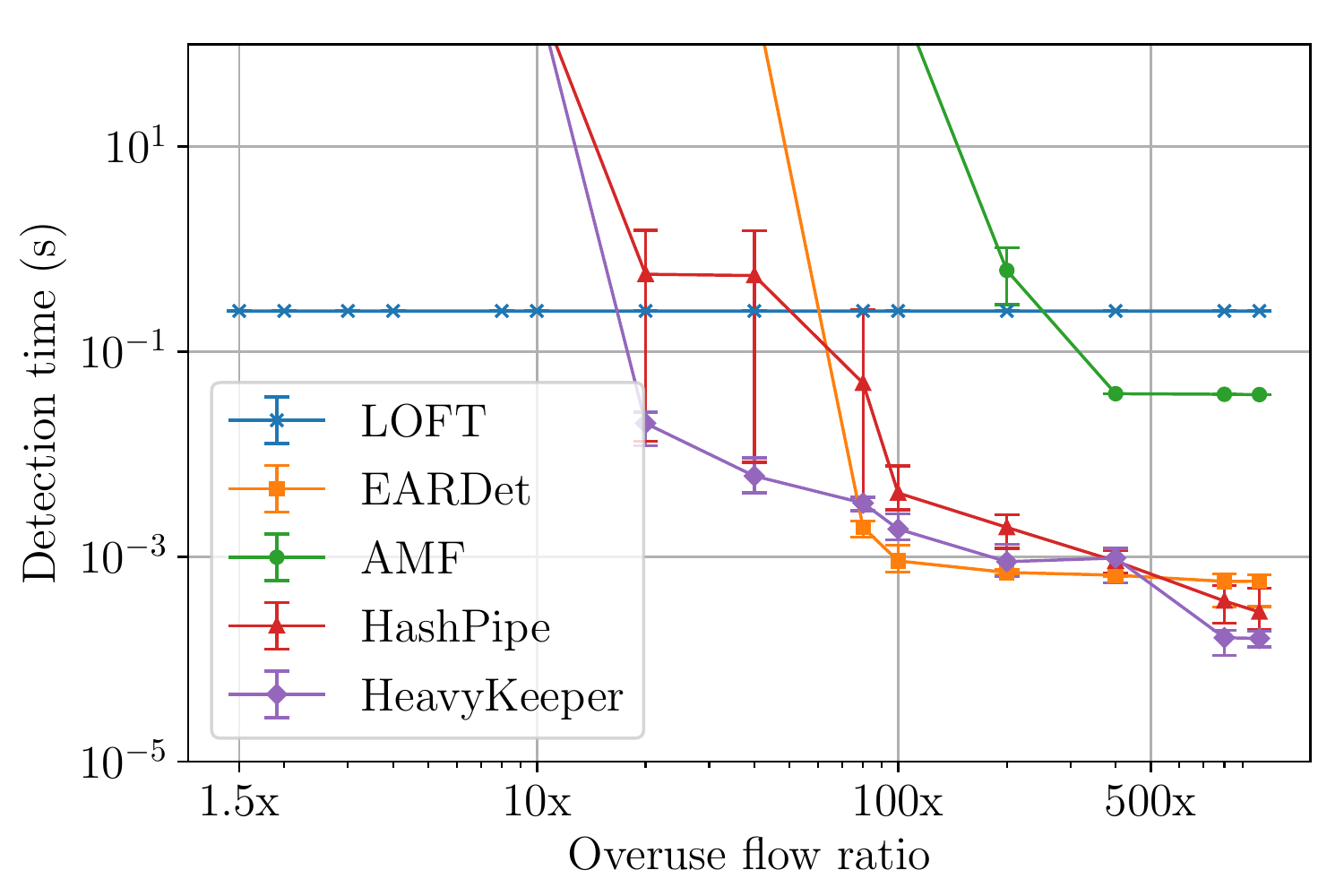}
		\label{fig:graph_cmp:10mio}
	}
    \caption{Detection delay of overuse flow detectors under different
        overuse ratios and flow numbers~$N$. The error bars show the minimum and maximum values over 100 runs.
    }
	\vspace*{-3pt}
\end{figure}

Figure~\ref{fig:graph_cmp} shows the detection delays of \name, EARDet,
HashPipe, HeavyKeeper and AMF under different overuse ratios $\ratio$ on a
log-log scale. Each data point is averaged over \num{100} runs. We find that
\name detects a \num{1.5}x overuse flow in less than one second, whereas all
other detectors fail to detect it before the \SI{300}{\second} timeout. For
larger overuse flows, \name still outperforms AMF, EARDet, HashPipe and
HeavyKeeper when the overuse ratio is less than \num{400}x, \num{7}x and
\num{3}x, respectively. As opposed to HashPipe and HeavyKeeper, \name delivers
high accuracy at a lower variance. Moreover, \name can achieve much higher
throughput (cf. \S\ref{sec:scalability}).

The reason that \name is slower in detecting high-rate overuse flows is that it
needs at least one major cycle, which takes \SI{0.25}{\second}, to select the
overuse flow. These results confirm that \name can efficiently detect overuse
flows using a small amount of router resources. While heavy-hitter detection
schemes perform better than \name regarding the extremely large flows that they
are designed to catch, these schemes are ineffective in low-rate overuse flow
detection, which is the goal of this paper.

As Figure~\ref{fig:graph_cmp:10mio} shows, these results are confirmed by
performing the same experiments with \num{10} million flows, which is the number
of flows to be expected on a \si{\tbps} link (cf.
Section~\ref{sec:definition:flowpolicingmodel}). For \num{10} million flows, the
higher accuracy of \name is even more prominent, as even the best other schemes
(i.e., HashPipe and HeavyKeeper) fail to detect the overuse flows for all
overuse ratios below \num{20}.

\subsection{Comparison: CAIDA Traffic Trace}
In addition to using synthesized background traffic in which every non-overuse
flow fully utilizes the reserved bandwidth, we also compare detectors on an
OC192 link using a real traffic trace, namely the CAIDA New York Anonymized
Internet Trace~\cite{caida201810}.
In the trace, the majority of the flows are smaller than \num{512} bytes per
second, and \num{95}\% of the flows are smaller than \num{14000} bytes per
second.


We first regulate every flow in the CAIDA trace with the permitted bandwidth
$\fsgamma$. Flows that use more than the permitted bandwidth are governed by
dropping overuse packets. Then, a 2-fold overuse flow is added into the traffic
trace. This setting reflects the scenario where an ISP wants to mitigate DDoS by
putting a bandwidth cap on individual flows. Because the number of flows in the
CAIDA traffic is smaller than in the fully utilized traffic trace, on this
smaller network, we only allocated \num{2048} + \num{64} counters to all detectors
(\num{64} flow monitors). \name is configured to run \num{4} slices of the
\spart per second.

\begin{figure}[t]
    \centering
    \vspace*{-3pt}
    \includegraphics[width=0.85\columnwidth]{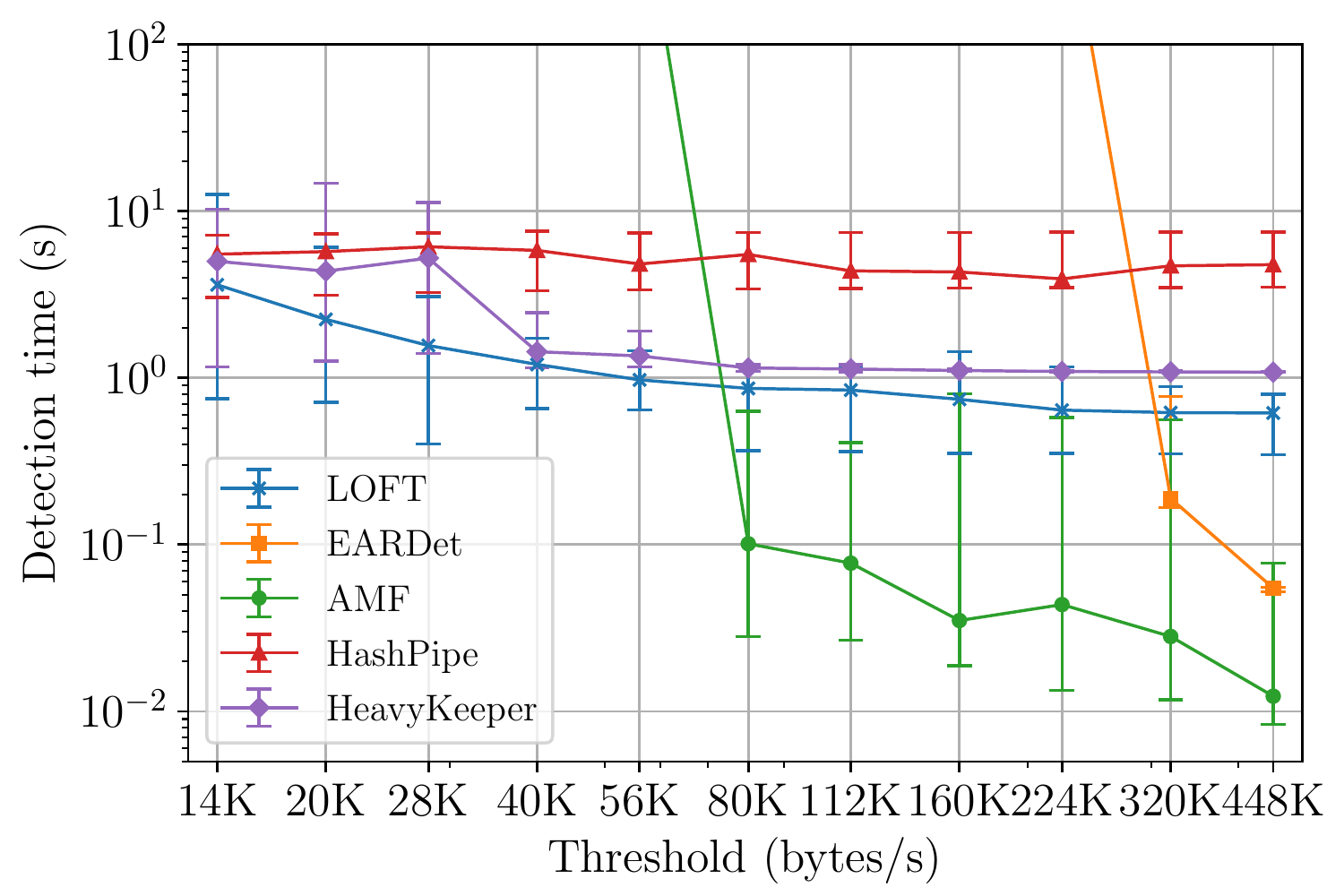}
    \vspace*{-8pt}
    \caption{Detection time of overuse flow detectors to catch a overuse flow sending $2\fsgamma$ given CAIDA traffic with different regulation threshold $\fsgamma$ and $\fsbeta = 1500$ for all.}
    \label{fig:graph_caida_cmp}
    \vspace*{-8pt}
\end{figure}

Figure~\ref{fig:graph_caida_cmp} shows the detection delays of \name, EARDet,
AMF, HashPipe and HeavyKeeper under different permitted bandwidths on a log-log
scale. Each data point is averaged over \num{100} runs. \name detects the 2-fold
overuse flow in \SIrange{0.3}{1.5}{\second}. EARDet and AMF can quickly and
reliably detect the 2-fold overuse flow only when it is much larger than the
majority of the non-overuse flows. HashPipe and HeavyKeeper catch the overuse
flow also given a low threshold, but these schemes still require a multiple of
\name's detection time to identify the overuse flow. Furthermore, the throughput
of these schemes is substantially lower than the throughput of \name (cf.
\S\ref{sec:scalability}).

\subsection{\name Sensitivity Tests}
\label{sec:sensitivity}
We have shown that given the same memory budget, \name can detect low-rate
overuse flows much faster than AMF, EARDet, HashPipe and HeavyKeeper. The
following experiments further investigate \name's effectiveness by varying its
parameters under a background traffic setting that is challenging for all
evaluated detectors.

\name is configured to run one slice of the \spart per second since it has to
process more flows in the following tests.  We consider a half-utilization
scenario, in which a half of the non-overuse flows send up to the permitted
bandwidth, and the rest sends almost negligible traffic. This is a more
challenging scenario than full utilization for \name because the variance
between counters are not only affected by the number of flows aggregated but
also by the variance of flow sizes.  This half-utilization scenario can capture
the behavior of typical streaming traffic, in which one direction is used to
send data and the other to send ACKs.

Half of the non-overuse flows send traffic up to the flow specification
$\fsgamma = \frac{400Gbps}{\flownum}$ and $\fsbeta = 1500$ and the other half
send 25 times less traffic. As before, one overuse flow is injected in the
simulation. This $\ratio$-fold overuse flow follows a flow specification
$\fsgamma = \frac{400Gbps}{\flownum} \cdot \ratio$ and $\fsbeta = 1500$. Each
data point of detection delay is the average of 100 simulations.

\begin{figure}[t]
    \centering
    \includegraphics[width=0.85\columnwidth]{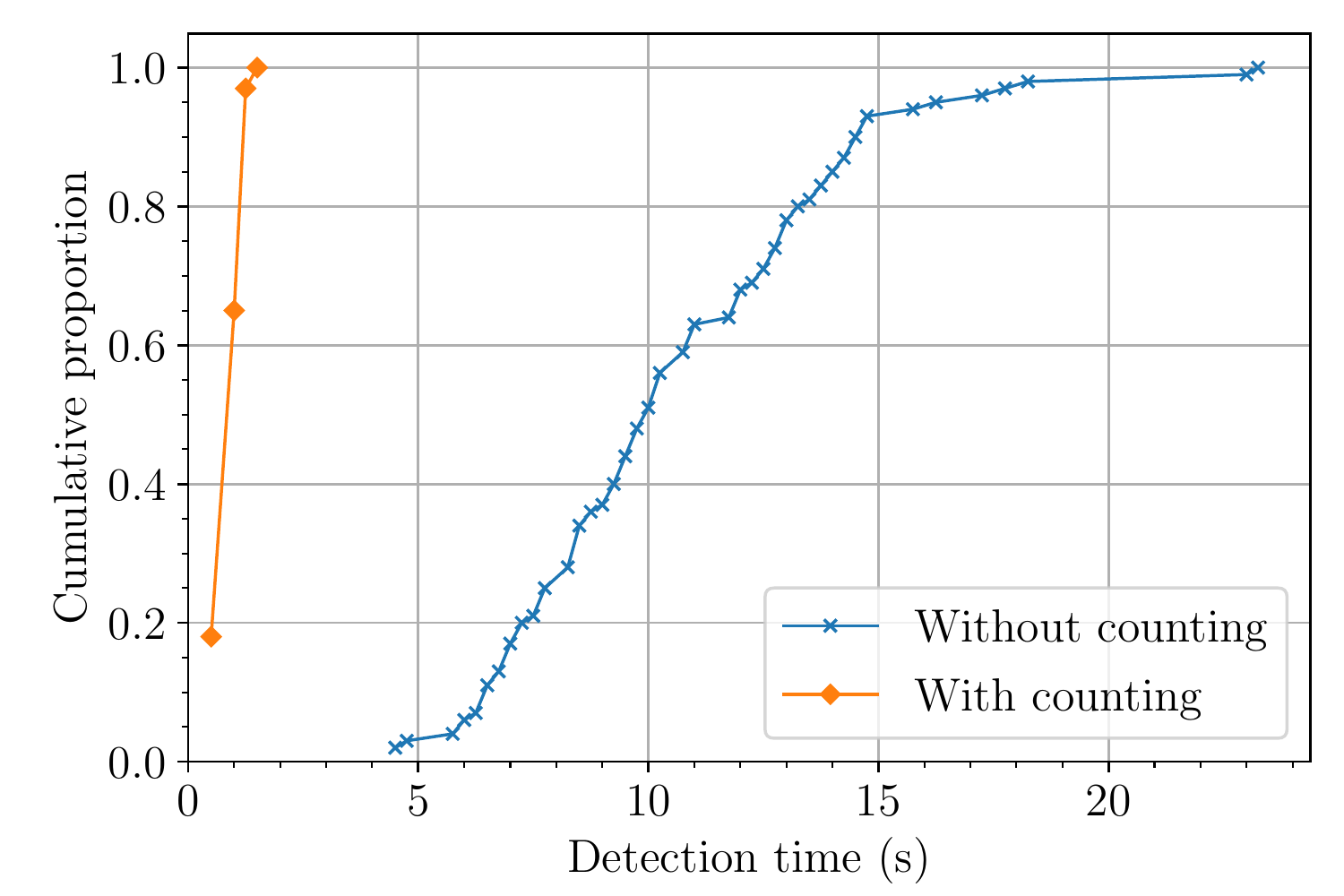}
    \vspace*{-8pt}
    \caption{\name's detection time of an overuse flow with and w/o counting. $\flownum = 130K$, $\fwidth = 16384$, $\ratio = 1.5$.}
    \label{fig:graph_woc}
    \vspace*{-8pt}
\end{figure}

\paragraph{Flow counting drastically reduces the detection time}
Figure~\ref{fig:graph_woc} shows that the estimator with flow counting and
dividing significantly outperforms the one without counting. This result
supports our perspective in Section~\ref{ssec:alg:overview} that the detection
accuracy of sketches is suffered by not taking the number of flows into account. In other words, \name is highly accurate thanks
to its efforts to reduce the counter noise that
stems from the variance of counter cardinality.

%

\begin{figure*}[t]
    \centering
    \subfigure[$\flownum = 400K$, $r = 0\%$, varying $\fwidth$]{
        \includegraphics[width=0.30\textwidth]{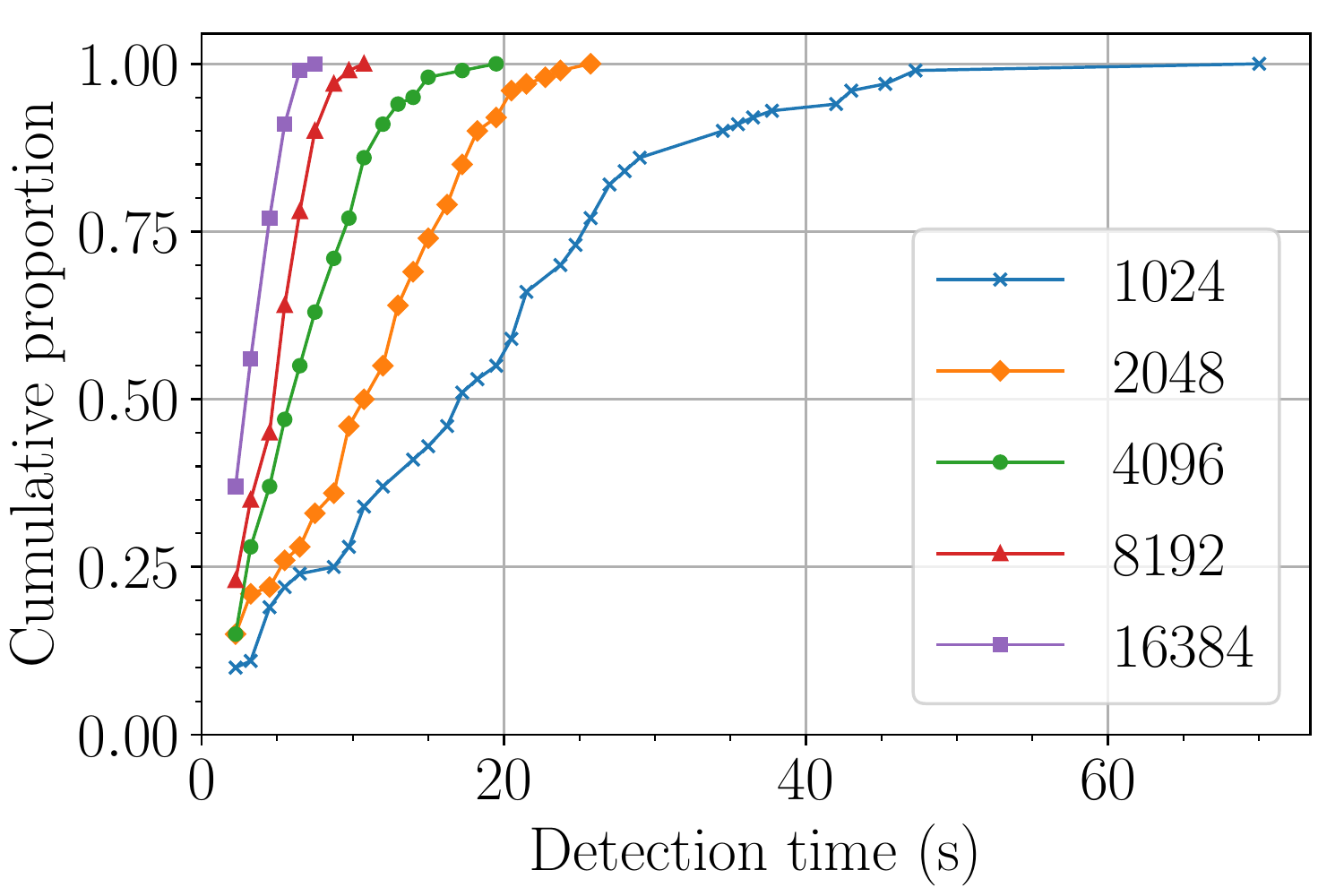}
        \label{fig:graph_counter}
    }
    \subfigure[$\fwidth = 1024$, $r = 0\%$, varying $\flownum$]{
        \includegraphics[width=0.30\textwidth]{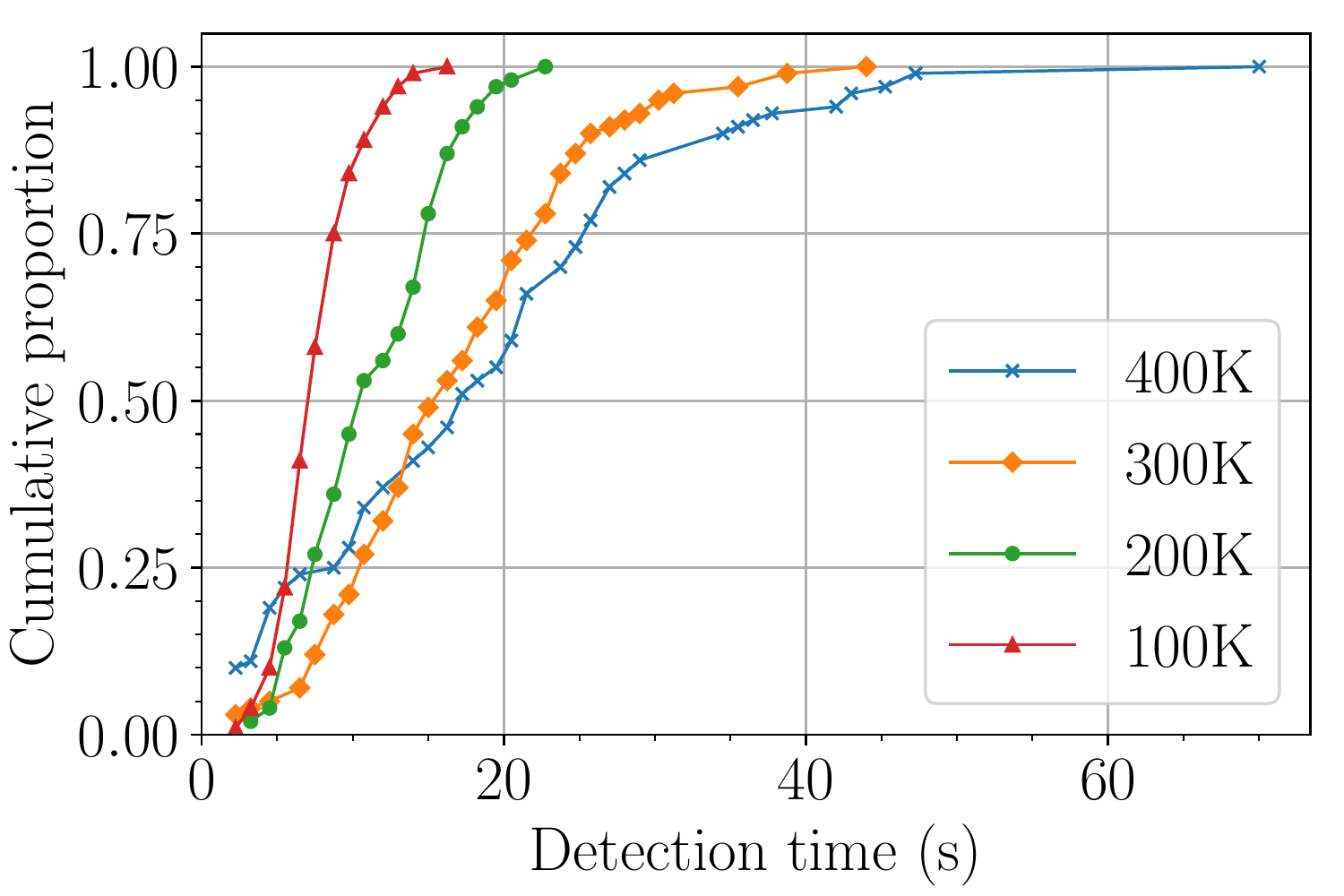}
        \label{fig:graph_num}
    }
	\centering
	\subfigure[$\flownum = 400K$, $\fwidth = 16,384$, varying $r$.]{
		\includegraphics[width=0.30\textwidth]{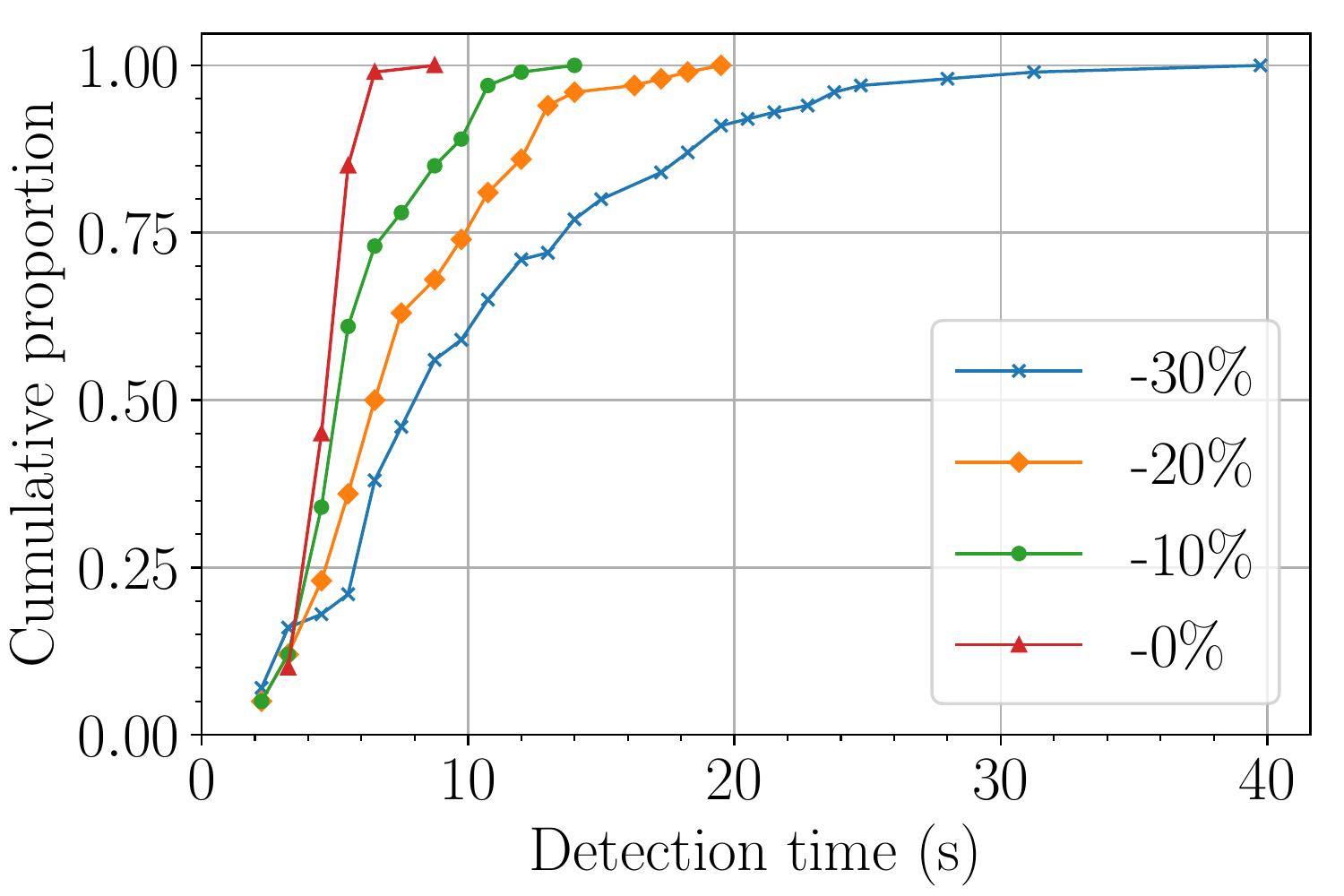}
		\label{fig:graph_miss}
	}
\vspace*{-5pt}
	\caption{CDF of \name's detection delay given
		number of counters ($\fwidth$), number of flows ($\flownum$), and
		sampling miss rate ($r$).}
    \vspace*{-10pt}
\end{figure*}

\paragraph{Memory budget}
In this experiment, we investigate the impact of fast memory size on the
detection delay. Figure~\ref{fig:graph_counter} shows the cumulative
distribution of the detection delay given different numbers of fast-memory
counters, ranging from \num{1024} to \num{16384}. As the number of fast-memory
counters is doubled, \name's detection speed increases as lowering the number of
flows sharing a counter reduces the variance of each counter.
However, even with the maximum number of counters and including the memory
required for monitoring suspicious flows, the fast-memory consumption of \name
is around \SI{130}{\kB}, which represents more than an order of magnitude
reduction compared to the \SI{3}{\MB} of fast memory needed for individual-flow
monitoring under the same traffic conditions (cf. \S
\ref{sec:definition:flowpolicingmodel}).


\paragraph{Number of non-overuse flows}
We evaluate the impact of the number of non-overuse flows on the detection
delay. Using \num{1024} fast-memory counters and one 2-fold overuse flow,
Figure~\ref{fig:graph_num} shows the cumulative detection delay given different
numbers of non-overuse flows, ranging from \num{100000} to \num{400000}. The
detection delay grows with the number of non-overuse flows because the variance
of counter cardinality increases as the total number of flows grows.

\paragraph{Imprecise active-flow list}
In our previous experiments, the fixed sampling rate (as defined in
Table~\ref{default-param}) is sufficient to maintain a precise active-flow list.
Given that maintaining this list is bound by the hardware-limited sampling rate,
the number of active flows might be too high to build a precise active-flow
list. For example, for \num{400000} active flows and a sampling rate of
\num{800000} flows per major cycle, the active-flow list will miss about
\num{15}\% of active flows.

To understand the impact of an imprecise active flows list on \name, we evaluate
different miss rates of active-flow lists. Figure~\ref{fig:graph_miss} shows
that in case of the half-utilization scenario, \name performs worse with
increasing imprecision of the active-flow list. This is due to the reason that
\name will use an inaccurate number of flows to calculate estimators, which
leads to higher variance. Nevertheless, even missing \num{20}\% of active flows,
\name can still catch the overuse flow under \SI{14}{\second} with \num{95}\%
probability.

\subsection{Scalability of \name}
\label{sec:scalability}

\subsubsection{DPDK Implementation}
\begin{figure}[htb]
\centering
\subfigure[Throughput given number of logical cores and packet
sizes.]{\includegraphics[trim=0 10 0 10, ,clip,width=0.85\columnwidth]{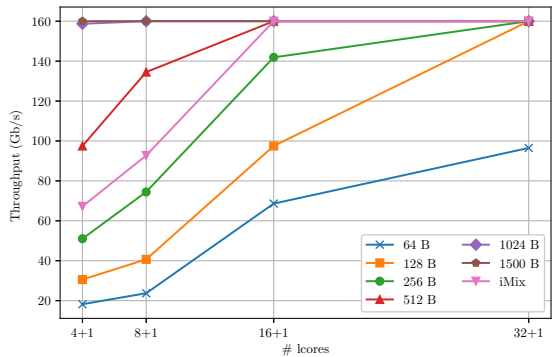}\label{fig:scalability}}
\subfigure[Overhead compared to regular L3 forwarding for 8+1 cores.\vspace{-18pt}]{\includegraphics[trim=0 10 0 10, ,clip,width=0.85\columnwidth]{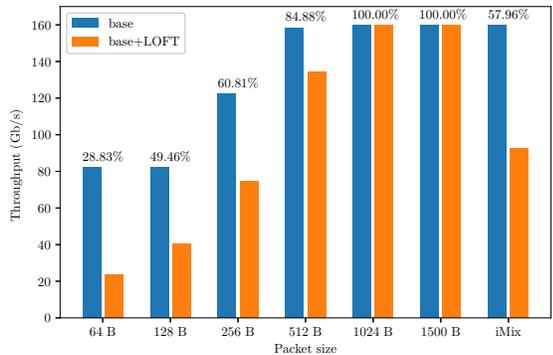}\label{fig:overhead}}
\caption{DPDK implementation results.}
\vspace*{-8pt}
\end{figure}

To understand the scalability of \name in a DPDK environment, we evaluate the
maximum packet rate with respect to the packet size and number of cores that
execute the \fpart concurrently. Figure~\ref{fig:scalability} shows that \name
is able to achieve line-rate for iMix-distributed traffic using \num{16} cores that
execute the \fpart and one core that runs the \spart. For smaller numbers of
cores, line-rate can only be achieved for larger packet sizes (\SI{1024}{\B}).

Since traffic flows with small sized packets perform considerably worse than
flows with large packet sizes, we additionally evaluate the overhead introduced
by \name by comparing it to regular L3 packet forwarding in DPDK.
Figure~\ref{fig:overhead} shows the throughput of regular L3 forwarding and the
throughput of forwarding with additional \name processing for different packet
sizes. Moreover, the figure gives the \name throughput as a percentage of the
corresponding base throughput. As visible in the figure, even regular L3 packet
forwarding using eight processing cores cannot achieve line-rate for packet
sizes smaller than \SI{1024}{\B}. Compared to regular packet forwarding, \name
introduces overhead for small packet sizes, which results in a maximum packet
rate of \textasciitilde{}\num{50} million packets per second (\si{\mpps}) using
eight processing cores, i.e., \textasciitilde{}\SI{6}{\mpps} per core. As for
larger packet sizes this maximum packet rate does not get exhausted, this effect
is diminished. However, \name still achieves a much higher packet rate than the
alternative schemes with the best accuracy, i.e., HashPipe and HeavyKeeper:
Prior work has shown a packet rate of \textasciitilde{}\SI{2}{\mpps} per core
for HashPipe~\cite{yang2018elastic} and a packet rate of
\textasciitilde{}\SI{2.5}{\mpps} per core for
HeavyKeeper~\cite{yang2019heavykeeper}.
%

\subsubsection{FPGA Implementation}
\label{sec:scalability:fpga}

We also implemented the fast-path component of \name (i.e., the \fpart using
\num{16384} counters) on a Xilinx Virtex UltraScale+ FPGA on the Netcope NFB
200G2QL platform~\cite{netcope-fpga} with two \SI{100}{\gbps} NICs and an
operating frequency of \SI{200}{\mega\hertz}. The \name implementation can
process a packet in every cycle. For minimum-size packets of \SI{64}{\B}, each
NIC manages to transfer one packet per cycle to the \name implementation, which
allows to achieve a packet rate of \SI{200}{\mpps} per NIC. As the FPGA platform
contains two NICs, it achieves a total packet rate of
\textasciitilde{}\SI{400}{\mpps}. This high throughput demonstrates that \name
is suitable for high-speed packet processing if implemented on programmable
NICs. The full implementation is described in a paper submitted to a specialized
conference~\cite{anonymous2020fpga}, as the FPGA-specific implementation details represent a contribution 
that goes beyond the scope of this paper.

\section{Theoretical Analysis Results}
\label{sec:damage}
\label{ssec:analysis:damage}

\newcommand{\resetcycleduration}{\ensuremath{T_{\mathit{reset}}}}


In the mathematical analysis provided 
in~\cref{sec:analysis},
we analyze
the overuse flow detection, and derive a lower bound 
(Equation~\ref{win_bound})
for the probability that an overuse flow is detected within a reset cycle, i.e.,
before a reset. This bound depends on a number of parameters of \name. The most
important are the duration of a reset cycle $\resetcycleduration$ (where
$\resetcycleduration \triangleq \parttime \resetcycle$) and the number of
fast-memory counters~$\fwidth$.

By requiring that the lower bound be equal to a desired detection probability,
we compute an upper bound on the required duration $\resetcycleduration$ of a
reset cycle for a given number of counters in order to achieve that detection
probability.

Table~\ref{tbl:graph_counter} shows the reset cycle duration
$\resetcycleduration$ calculated under the settings from the experiment of
Figure~\ref{fig:graph_counter}, compared to the real detection delays as
obtained by our analysis in Section~\ref{sec:sensitivity}. The result shows that
our analysis does not underestimate the detection delay, since the upper bound
on the reset cycle duration, which is the worst-case-estimate of the 95th
percentile of the detection delay, is always higher than those obtained by our
evaluation. The reasons why the estimated delay is \numrange{2}{5} higher than
the real one are twofold. First, the synthesized traffic in our experiments does
not always represent the worst-case scenario. Second, \name ranks and detects
overuse flows at the end of every \spart, but our analysis ignores the
probability that overuse flows might be caught at some point before
$\resetcycle$, which loosens the bound.

\begin{table}[h]
    \centering
    \caption{Calculated reset cycle duration ($\resetcycleduration$) and the
        real 95th percentile of the detection delay from our evaluation with
        different numbers of counters in fast memory in the experiment of
        Figure~\ref{fig:graph_counter}.
    }
    \vspace*{5pt}
    \label{tbl:graph_counter}
    \resizebox{\columnwidth}{!}{%
        \begin{tabular}{lrrrrr}
            \toprule
            \makecell[l]{Num.\@ of counters}  & \num{1024}  & \num{2048}  & \num{4096}  & \num{8192} & \num{16384} \\
            \midrule
            Reset cycle (\si{\second})                   & \num{231}   & \num{116}  & \num{58}    & \num{29}   & \num{15}    \\
            \makecell[l]{Detection delay (\si{\second})} & \num{43.03} & \num{20.48} & \num{14.04} & \num{8.67} & \num{6.52}  \\
            \bottomrule
        \end{tabular}%
    }
    \vspace*{-10pt}
\end{table}

\section{Related Work}
\label{sec:related}
Despite a significant amount of research on the problem of
detecting high-rate overuse flows, the problem of efficiently detecting low-rate
overuse flows has so far been neglected. A number of related schemes have been proposed to address similar problems, such
as large flow detection and top-k flow detection. Their ideas might be
applicable to detecting overuse flows. However, we discuss in the following
paragraphs that they are either orthogonal to our research direction, or that
they are insufficient to solve our problem as stated in
Section~\ref{sec:definition}.

The problem of detecting top-k flows aims to identify $k$ flows that consume
most of a link's bandwidth. A recent proposal called
HashPipe~\cite{sivaraman2017heavy} tackles the heavy hitter detection problem on
programmable hardware. To ensure line-rate detection given a limited amount of
fast memory, HashPipe constructs a pipeline of hash tables to efficiently
implement the Space Saving algorithm~\cite{metwally2005efficient}, such that the
heavier flows are more likely to be kept in the next stage of the pipeline.
However, our results in \S\ref{sec:evaluation} show that HashPipe fails to
detect the overusing flow in cases where the difference between overuse and
non-overuse flow is low (i.e., with an overuse ratio of \numrange{1.5}{2}). As
another top-k detection scheme, HeavyKeeper \cite{yang2019heavykeeper}, seems to
suffer from the same weakness, these systems seem unable to effectively filter
out the counter noise.



Similar to top-k flow detection, large-flow detection algorithms can be applied
to detecting overuse flows. Large flow detection algorithms identify flows that
use more than a threshold amount of bandwidth, and it is common that the higher
the threshold the better the performance will be. Besides AMF (introduced in
Section~\ref{ssec:definition:comparison}), CLEF~\cite{Wu2018} proposes to detect
low-rate overuse flows, which are similar to the low-rate overuse flows in our
work, using recursive division and by combining two detectors with complementing
properties. Our evaluation shows that \name outperforms EARDet, one of the
detectors used by CLEF, when the overuse ratio is lower than \num{7}x. The other
detector used by CLEF is a sketch and thus inherits the limitations explained in
Section~\ref{ssec:definition:comparison}.

Hybrid SRAM/DRAM-based architectures for exact counting have been proposed by
Shah \etal~\cite{shah2001analysis} and were further improved by Ramabhadran and
Varghese~\cite{ramabhadran2003efficient} and by Zhao
\etal~\cite{zhao2006design}. By default, these schemes only consider the number
of packets per flow, but not the flow sizes. Even though the authors propose
an extension to consider flow sizes, the use of probabilistic counting in these
schemes introduces a high counter variance, making low-rate overuse flows hard
to detect.
Lall \etal~\cite{Lall2009} propose another SRAM/DRAM hybrid data structure for
efficient detection of both medium and large flows. However, because the
proposed flow monitoring solution uses shared counters (in the form of spectral
Bloom filters), it performs poorly in catching low-rate overuse flows.


Another approach to tackle aforementioned problems uses \emph{sampling}, where
only a small subset of packets is used for flow accounting. Sampled
NetFlow~\cite{Netflow} is a widely deployed solution that collects one out of
every $n$ packets, and estimates statistics of the original population by
extrapolating from the sample. Researchers have proposed advanced sampling
algorithms tailored for catching large flows. For example, Sample and
Hold~\cite{Estan03} and Sticky Sampling~\cite{Manku2002} are designed to bias
toward large flows. Instead of using a static sampling rate, several adaptive
sampling algorithms dynamically adjust the sampling rate so as to keep resource
consumption under a fixed memory limitation~\cite{Estan2004,
Sanjuas-Cuxart2012}. However, without a sufficiently high sampling rate
(resulting in a large amount of fast memory), sampling-based algorithms are
prone to false positives and false negatives, as shown by Estan and
Varghese~\cite{Estan03}.
\name relies on sampling to generate a list of active flows (if the list is not
provided). However, \name ensures that at least one packet appears in a sample,
thus requiring a lower sampling rate than for accurate flow accounting.

A recent series of work including SketchVisor~\cite{huang2017sketchvisor},
ElasticSketch~\cite{yang2018elastic} and NitroSketch~\cite{liu2019nitrosketch}
devises techniques to speed up the updating of detector datastructures based on
sketches. However, these techniques all trade off detection accuracy against
processing speed, i.e., these algorithms achieve even lower accuracy than the
unaltered sketches like AMF evaluated in Section~\ref{sec:evaluation}.
In contrast, \name can achieve high accuracy \emph{and}
a low per-packet overhead.

\section{Conclusions}
\label{sec:conclusion}

Due to limitations of previous approaches to probabilistic flow monitoring,
network operators so far lacked effective measurement tools that could give an
accurate insight into the flow-size distribution on high-capacity routers. Given
router constraints regarding fast memory and computation, existing schemes
suffer from a large measurement error, which only allows the reliable detection
of extremely large flows, but not flows with a small amount of overuse. In this
work, we show that the source of this measurement error in sketch-based schemes
is \emph{counter noise}, i.e., the high variance of counter values. Using this
insight, we develop \name, a sketch-based approach that counteracts the counter
noise while respecting the stringent complexity constraints of high-speed
routers.

As a result, the measurement error of \name is so small that low-rate overuse
flows (i.e., flows only \numrange{50}{100}\% larger than the average flow) can
be reliably detected with a small amount of fast memory. Concretely, \name can
reliably identify a flow that is only \num{50}\% larger than the average flow
within one second, whereas all other investigated schemes fail to identify such
a flow even within \SI{300}{\second}. Moreover, \name accomplishes such high
accuracy while reducing the fast-memory requirement by more than one order of
magnitude in comparison with individual-flow monitoring. We also investigate
scalability and overhead of \name with a DPDK and an FPGA implementation, and
show that \name enables line-rate forwarding of a realistic traffic mix.

With these demonstrated properties, \name can serve as a powerful
flow-monitoring tool, which will allow network operators to improve the efficacy
of existing applications based on flow-size estimation (e.g., flow-size aware
routing) and to enable new applications based on such estimates (e.g.,
reservation-based DDoS defense).

\bibliographystyle{plain}
\bibliography{ESCALATE}

\appendix

\section{Mathematical Analysis}
\label{sec:analysis}

We would like to know the probability that \name can catch overuse
flows and how much damage they can do before being caught.  Our
analysis first derives a lower bound on the probability that a
non-overuse flow estimate becomes larger than an overuse flow
estimate. Then we use this bound to calculate the probability that an
overuse flow estimate falls within the top-$\fmwidth$ largest
estimates, which is also the probability of being monitored.

Due to the complexity of this problem, we make the following
assumptions throughout the analysis.
We require that the number of flows $\flownum$
in traffic is fixed during each detection period and $\flownum$ is large
enough so that the attacker is unable to manipulate either $\flownum$ or
the overall traffic size distribution. In addition, we assume that all
flow estimates, which are random variables in our model, only have
weak pairwise dependence, and see them as i.i.d. in the following
analysis.


\subsection{Lower Bound of Detection Probability}
\label{ssec:analysis:probability}

First we derive a lower bound of the dectection probability given the amount of
an overuse flow after \name starts a reset cycle. This lower bound helps us
analyze how long \name will need to catch the overuse flow with high enough
probability. Then by setting a proper reset cycle, our analysis determines the
maximum damage an overuse flow can do, and guarantees that \name will have high
probability to catch the overuse flow if it exceeds the limit of maximum
damage.

The detection probability of an overuse flow is exactly the probability of
\name choosing to monitor that flow and the flow monitor catches its overusing
behavior. Here we ignore the time spent in the flow monitor when an overuse
flow is reported to it. Analyzing and improving flow monitor is outside the
scope of this paper. Therefore, the detection probability is equal to the
probability of the overuse flow being selected by \name.

We define $\estim_i$, $\accum_i$, $\counter_i$ as the estimate,
accumulator, and counter of flow $i$ in our algorithm,
respectively. Hence they have the following relationship:
\begin{equation}
    \estim_i = \frac{\accum_i}{\counter_i}
\end{equation}

In each minor cycle, each flow, except for flow $i$, has $\frac{1}{\fwidth}$
chance to contribute part of its flow size to $\accum_i$. We define a
\emph{flow segment} $\flowseg_j$ as the amount contributed by another flow $j$
to $\accum_i$ in a minor cycle. Every $\flowseg_j$ can be seen as a random
variable with unknown distribution, since it is selected randomly by a uniform
hash function. More deeply, suppose $S_a = \{ \flowseg_1, \flowseg_2, \cdots,
    \flowseg_a \}$ and $S_b = \{ \flowseg_{a + 1}, \flowseg_{a + 2}, \cdots,
    \flowseg_{a + b} \}$ are added to $\accum_i$ in minor cycle $x$ and $x + 1$,
respectively. $S_a$ are exactly the $a$ uniform samples without replacement
from $\flownum$ flow segments sent by $\flownum$ flows during minor cycle $x$.
In addition, $S_b$ are another $b$ uniform samples from minor cycle $x + 1$ and
being independent of the prior samples $S_a$, since $S_a$ and $S_b$ are sampled
independently from two different sets of flow segments. These properties allow
us to apply Hoeffding's inequality~\cite{hoeffding1963} and derive the bound in
Equation~\ref{bound_inq}.

Because we assume that overuse flows are minority and cannot affect the traffic
distribution, and because all other flows will not violate the flow
specification, $\flowseg_j$ also satisfies the following bound, where $\boundv
    \triangleq \fsgamma \parttime + \fsbeta$.
\begin{equation} \label{seg_bound}
    \begin{aligned}
        0 \leq \flowseg_j, E[\flowseg_j] \leq \boundv
    \end{aligned}
\end{equation}

Suppose \name has run $\parts$ minor cycles, given there are $\rcounter_i$ flow
segments accumulated in $\counter_i$, the size of flow $i$ is $\flowsize_i$, we can represent $\estim_i$ as a random
variable under this condition as below.
\begin{equation} \label{basic_form}
    \estim_i|(\counter_i = \rcounter_i) = \frac{1}{\rcounter_i}(\flowsize_i + \sum_{j=1}^{\rcounter_i - \parts} \flowseg_j)
\end{equation}

From Equation~\ref{basic_form}, now $\estim_i$ is modeled by the sum
of several random variables, which is also a random variable with
unknown distribution. As mentioned previously, different $\estim$s are
seen as i.i.d.  This assumption is reasonable because under random
dispatching the probability that any pair of $\estim$ share many
$\flowseg$ (and thus build strong dependence) is extremely low.
The more precise analysis is left for future work.

Let $\estim_\llabel$ and $\estim_\blabel$ be the estimates of an overuse flow
and a non-overuse flow respectively. By Equation~\ref{basic_form}, we can enumerate
all conditions and derive the equation below (In the following equations, we
short $\counter_i = \rcounter_i$ to $\overline{\rcounter_i}$).
\begin{equation} \label{comp_form}
    P(\estim_\llabel > \estim_\blabel) =
    \sum_{\rcounter_\llabel=1}^{\flownum \parts} \sum_{\rcounter_\blabel=1}^{\flownum \parts}
    P(\estim_\llabel > \estim_\blabel|\overline{\rcounter_\llabel}, \overline{\rcounter_\blabel})
    P(\overline{\rcounter_\llabel}, \overline{\rcounter_\blabel})
\end{equation}

Instead of bounding Equation~\ref{comp_form} directly, we choose a threshold $\midp$ and simplify the bound with the following inequality, which holds since $\estim_\llabel$ and $\estim_\blabel$ are i.i.d. in our assumption.
\begin{equation}
    P(\estim_\llabel > \estim_\blabel|\overline{\rcounter_\llabel}, \overline{\rcounter_\blabel}) \geq
    P(\estim_\llabel > \midp|\overline{\rcounter_\llabel})P(\estim_\blabel < \midp|\overline{\rcounter_\blabel})
\end{equation}

Combining with Equation~\ref{comp_form}, we have the following inequality.
\begin{equation}
    P(\estim_\llabel > \estim_\blabel) \geq
    \sum_{\rcounter_\llabel=1}^{\flownum \parts} \sum_{\rcounter_\blabel=1}^{\flownum \parts}P(\estim_\llabel > \midp|\overline{\rcounter_\llabel})P(\estim_\blabel < \midp|\overline{\rcounter_\blabel})
    P(\overline{\rcounter_\llabel}, \overline{\rcounter_\blabel})
\end{equation}

Now the goal is deriving the bound of $P(\estim_i >
    \midp|\overline{\rcounter_i})$. Suppose $\rcounter_\blabel - \parts \geq 1$, for the estimate of
non-overuse flow $\estim_\blabel$, combining with Equation~\ref{basic_form}, we have
the following inequality.
\begin{equation} \label{ori_inq}
    \begin{aligned}
        P & (\estim_\blabel \geq \midp|\overline{\rcounter_\blabel})                                                                                                                                                                          \\
          & = P(\frac{1}{\rcounter_\blabel}(\flowsize_\blabel + \sum_{j=1}^{\rcounter_\blabel - \parts} \flowseg_j) \geq \midp)                                                                                                               \\
          & = P(\frac{1}{\rcounter_\blabel}\flowsize_\blabel + \frac{\rcounter_\blabel - \parts}{\rcounter_\blabel}\frac{1}{\rcounter_\blabel - \parts}\sum_{j=1}^{\rcounter_\blabel - \parts} \flowseg_j \geq \midp)                         \\
          & = P(\frac{1}{\rcounter_\blabel - \parts}\sum_{j=1}^{\rcounter_\blabel - \parts} \flowseg_j \geq \frac{\rcounter_\blabel}{\rcounter_\blabel - \parts}(\midp - \frac{\flowsize_\blabel}{\rcounter_\blabel}))                        \\
          & = P((\frac{1}{\rcounter_\blabel - \parts}\sum_{j=1}^{\rcounter_\blabel - \parts} \flowseg_j) - E[\flowseg] \geq \frac{\rcounter_\blabel}{\rcounter_\blabel - \parts}(\midp - \frac{\flowsize_\blabel}{\rcounter_\blabel}) - E[X])
    \end{aligned}
\end{equation}

Let $v \triangleq \frac{\rcounter_\blabel}{\rcounter_\blabel - \parts}(\midp -
    \frac{\flowsize_\blabel}{\rcounter_\blabel}) - E[\flowseg]$. If $v \geq 0$,
since $\flowseg_j$ is constrainted by the inequality in
Equation~\ref{seg_bound} and satisfies sufficient properties as we mentioned
before, we can apply Hoeffding's inequality to get an upper bound of
Equation~\ref{ori_inq} as below.
\begin{equation} \label{bound_inq}
    \begin{aligned}
        P((\frac{1}{\rcounter_\blabel - \parts}\sum_{j=1}^{\rcounter_\blabel - \parts} \flowseg_j) - E[\flowseg] \geq v) \leq \exp(-\frac{2 (\rcounter_\blabel - \parts) v^2}{\boundv^2})
    \end{aligned}
\end{equation}

Here, we choose the midpoint between the expected values of
$\estim_\llabel|\overline{\rcounter_\llabel}$ and
$\estim_\blabel|\overline{\rcounter_\blabel}$ as $\midp$. Therefore, $\midp$
for $\overline{\rcounter_\llabel}, \overline{\rcounter_\blabel}$ can be
represented as below.
\begin{equation} \label{midp_eq}
    \begin{aligned}
        \midp_{\rcounter_\llabel \rcounter_\blabel}
         & = \frac{1}{2}(E[\estim_\llabel|\overline{\rcounter_\llabel}] + E[\estim_\blabel|\overline{\rcounter_\blabel}])                                                                                           \\
         & = E[\flowseg] + \frac{1}{2 \rcounter_\llabel \rcounter_\blabel}(-\parts E[\flowseg] (\rcounter_\llabel + \rcounter_\blabel) + \rcounter_\blabel \flowsize_\llabel + \rcounter_\llabel \flowsize_\blabel)
    \end{aligned}
\end{equation}

Finally, we substitute $\midp$ of inequality~\ref{bound_inq} by
Equation~\ref{midp_eq} and derive the following bound.
\begin{equation} \label{ori_bound_b}
    \begin{aligned}
        P(\estim_\blabel \geq \midp_{\rcounter_\llabel \rcounter_\blabel}|\overline{\rcounter_\blabel})
                               & \leq P((\frac{1}{\rcounter_\blabel - \parts}\sum_{j=1}^{\rcounter_\blabel - \parts} \flowseg_j) - E[\flowseg] \geq v_\blabel)                                                                               \\
                               & \leq \exp(-\frac{2 (\rcounter_\blabel - \parts) v_\blabel^2}{\boundv^2})                                                                                                                                    \\
        \text{where}~v_\blabel & \triangleq \frac{1}{2\rcounter_\llabel(\rcounter_\blabel - \parts)}[\parts E[\flowseg] (\rcounter_\llabel - \rcounter_\blabel) + \rcounter_\blabel \flowsize_\llabel - \rcounter_\llabel \flowsize_\blabel]
    \end{aligned}
\end{equation}

For the estimate of overuse flow $\estim_\llabel$, we can use the same method to derive its bound as below.
\begin{equation} \label{ori_bound_l}
    \begin{aligned}
        P(\estim_\llabel \leq \midp_{\rcounter_\llabel \rcounter_\blabel}|\overline{\rcounter_\llabel})
                               & \leq \exp(-\frac{2 (\rcounter_\llabel - \parts) v_\llabel^2}{\boundv^2})                                                                                                                                    \\
        \text{where}~v_\llabel & \triangleq \frac{1}{2\rcounter_\blabel(\rcounter_\llabel - \parts)}[\parts E[\flowseg] (\rcounter_\llabel - \rcounter_\blabel) + \rcounter_\blabel \flowsize_\llabel - \rcounter_\llabel \flowsize_\blabel]
    \end{aligned}
\end{equation}

Although Equation~\ref{ori_bound_b} and Equation~\ref{ori_bound_l} contain an
unknown variable $E[\flowseg]$ and the inequalities only hold under certain
conditions, since $0 \le E[\flowseg] \le M$, we can get the worst-cast probabilities by choosing $E[\flowseg]$ to minimize $v_\blabel$ and $v_\llabel$. Therefore, Equation~\ref{ori_bound_b} and Equation~\ref{ori_bound_l} can be rewritten into the following two functions, which give out the worst-case bounds without knowing $E[\flowseg]$ or making any assumption on $\rcounter_\blabel$ and $\rcounter_\llabel$.
\begin{equation}
    \begin{aligned}
         & \hat{P_\blabel}(\parts, \rcounter_\blabel, \rcounter_\llabel, \flowsize_\blabel, \flowsize_\llabel) \\
         & \triangleq
        \begin{cases}
            1                                                                           & \text{if } \min{v_\blabel} < 0 \lor \rcounter_\blabel - \parts < 1 \\
            \exp(-\frac{2 (\rcounter_\blabel - \parts) (\min{v_\blabel}^2)}{\boundv^2}) & \text{otherwise}
        \end{cases}
        \\
         & \hat{P_\llabel}(\parts, \rcounter_\blabel, \rcounter_\llabel, \flowsize_\blabel, \flowsize_\llabel) \\
         & \triangleq
        \begin{cases}
            1                                                                           & \text{if } \min{v_\llabel} < 0 \lor \rcounter_\llabel - \parts < 1 \\
            \exp(-\frac{2 (\rcounter_\llabel - \parts) (\min{v_\llabel}^2)}{\boundv^2}) & \text{otherwise}
        \end{cases}
    \end{aligned}
\end{equation}

For the pmf of $P(\overline{\rcounter_\llabel}, \overline{\rcounter_\blabel})$,
we assume that our random flow-to-counter mapping in each minor cycle makes
$\counter_i$ follow the binomial distribution, so it can be represented
as below.
\begin{equation}
    \begin{aligned}
        P(\parts, \overline{\rcounter_\llabel}, \overline{\rcounter_\blabel}) =
        P(B(\flownum \parts, \frac{1}{\fwidth}) = \rcounter_\llabel) P(B(\flownum \parts, \frac{1}{\fwidth}) = \rcounter_\blabel)
    \end{aligned}
\end{equation}

Combining all equations above, given the amounts of an overuse flow
$\flowsize_\llabel$ and a non-overuse flow $\flowsize_\blabel$ sent in $\parts$
minor cycles, we can derive the lower bound function $\hat{P}_{win}$ of $P(\estim_\llabel > \estim_\blabel)$ as below.
\begin{equation}
    \begin{aligned}
         & \hat{P}_{win}(\parts, \flowsize_\llabel, \flowsize_\blabel)                                          \\
         & \triangleq \sum_{\rcounter_\llabel=1}^{\flownum \parts} \sum_{\rcounter_\blabel=1}^{\flownum \parts}
        (1 - \hat{P_\blabel}(\parts, \rcounter_\blabel, \rcounter_\llabel, \flowsize_\blabel, \flowsize_\llabel))
        (1 - \hat{P_\llabel}(\ldots))
        P(\parts, \overline{\rcounter_\llabel}, \overline{\rcounter_\blabel})
    \end{aligned}
\end{equation}

The worst-case of $\hat{P}_{win}$ happens when the amount of the non-overuse flow $\flowsize_\blabel$ is equal to its maximum size. Therefore, $\hat{P}_{win}(\parts, \flowsize_\llabel, \fsgamma \parttime \parts + \fsbeta)$ gives out the worst-case lower bound when the behavior of non-overuse flows is unknown.

The probability that an overuse flow will be selected into the flow monitor is
equal to the probability that it loses to less than $\fmwidth$ non-overuse flows
during ranking. Therefore, the lower bound of this probability can be
represented as below, since we have assumed that $\estim$ are i.i.d.
\begin{equation} \label{win_bound}
    \begin{aligned}
        P_{mon}(\parts, \flowsize_\llabel) \geq \sum_{k=0}^{\fmwidth - 1} \binom{\flownum}{k} (1 - \widetilde{P}_{win})^k (\widetilde{P}_{win})^{\flownum - k} \\
        \text{where}~\widetilde{P}_{win} \triangleq \hat{P}_{win}(\parts, \flowsize_\llabel, \fsgamma \parttime \parts + \fsbeta)
    \end{aligned}
\end{equation}

We use this bound in Section~5 in the paper, showing
how its predictions match the results of our evaluation

\end{document}